\documentstyle[12pt]{article}
\def\pin{$\mbox{}$\indent}  
\def\erf{\mbox{\rm erf}}
\begin{document}
\setcounter{page}{0}
\begin{titlepage}
\title{A canonical ensemble approach to graded-response perceptrons}
\author{ D.~Boll\'e$^{(1)}$ and R. Erichsen Jr.$^{(2)}$\\
           Instituut voor Theoretische Fysica \\
            K.U.\ Leuven \\
             B-3001 Leuven, Belgium}
%
\date{}
\maketitle
\thispagestyle{empty}
\begin{abstract}
\normalsize
\noindent
Perceptrons with graded input-output relations and a limited output
precision are studied within the Gardner--Derrida canonical ensemble
approach. Soft non-negative error measures are introduced allowing for
extended retrieval properties. In particular, the performance of these
systems for a linear and quadratic error measure, corresponding to the
perceptron respectively the adaline learning algorithm, is compared with
the performance for a rigid error measure, simply counting the number of
errors. Replica--symmetry--breaking effects are evaluated.
\end{abstract}
\vspace*{1cm}
\newline \noindent
Short title: Graded perceptrons with soft error measures.
\newline \noindent
$^{(1)}$e-mail: Desire.Bolle@fys.kuleuven.ac.be
\newline \noindent
Also at Interdisciplinair Centrum voor Neurale  Netwerken, K.U.Leuven, Belgium
\newline \noindent
$^{(2)}$e-mail: rubem@if.ufrgs.br
\newline \noindent
Present address: Instituto de F\'{\i}sica, Universidade Federal do
Rio Grande do Sul,
\newline  \noindent 
Caixa Postal 15051, 91501--970 Porto Alegre, RS, Brazil
\end{titlepage}
\section*{1. Introduction}
\pin
Graded--response perceptrons constitute the basic building blocks of
layered architectures trained by the backpropagation algorithm. This
motivates the interest in these systems over the last years. Questions
pertaining to retrieval properties of specific
architectures \cite{T}-\cite{KB}, to optimal capacities of networks
designed to perform a given storage task \cite{BKvM}-\cite{BE} and to
generalisation abilities \cite{BKO} have been adressed by statistical
mechanics approaches.

In this paper, we develop a Gardner--Derrida (GD) type analysis \cite{GD}
of the optimal storage properties for graded--response
perceptrons when allowing errors. The underlying idea thereby is to view
learning in these perceptrons as an optimization process
in the space of couplings. By introducing soft non--negative error measures
we investigate the canonical ensemble
generated by the corresponding cost function in the space of couplings
using the replica method. In this discussion we allow for a limited output
precision in the storage task to be solved
by the perceptron. In particular, a linear and a quadratic error measure
are investigated. The corresponding cost functions are of special interest
since they
define a perceptron learning algorithm respectively an adaline learning
algorithm through the method of gradient descent.
For comparison we also derive the results for the rigid GD
error measure that simply counts the number of errors. Replica--symmetric
(RS) and first--step replica--symmetry--breaking (RSB) solutions for the
storage capacity and the average output error are studied.

For the case of two--state atractor neural networks the canonical ensemble
approach advocated in ref.~\cite{GD} has been streamlined and extended to
other cost functions than the rigid one \cite{GG}. The methods and results
obtained there are, of course, also relevant for perceptron networks.
First--step RSB effects above the critical capacity have then been studied
in \cite{ET} for binary
perceptron networks with a GD cost function and have been
extended to other cost functions \cite{MEZ}-\cite{WSW}. Recently, it
has been shown \cite{WS} for the GD
cost function that in the region above the critical capacity full RSB is
necessary for an exact solution. A direct evaluation of the two--step RSB
solution has been performed in this case, yielding a minimum storage error
only slightly greater than the
one-step RSB. The conclusion was put forward that for most practical
purposes one--step RSB will be adequate.

The rest of this paper is organized as follows. In section 2 we shortly
review the canonical approach adapted to the graded--response perceptron
and introduce the different cost functions we want to consider: the rigid
one, the linear one and the quadratic one. Section 3 contains the replica
theory for these cost functions and determines the critical storage
capacity, the distribution
of the local fields and the average output error. Both the RS approximation
and the first step RSB are treated for a general monotonic input--output
relation. Section 4 describes the results of this theory applied to two
specific, frequently used input--output relations, i.e., the hyperbolic
tangent and
the piecewise linear one. In section 5 the most important results are
summarized. Finally, the appendix contains the technical details of the
derivations.

The analysis reported on in this work extends our results
\cite{BKvM,BE} on the optimal capacity of graded--response perceptrons in
the framwework of the Gardner theory \cite{Gardner}.

\section*{2. Canonical ensemble approach}
\pin
The task to be solved by the graded--response perceptron is to map a
collection of input patterns $\{\xi_i^\mu; 1\leq i\leq N\}$, $1\leq
\mu\leq p$, onto a corresponding set of outputs $\zeta^\mu$, $1\leq
\mu\leq p$, via
\begin{eqnarray}
        \zeta^\mu & = & g\left(\gamma h^\mu \right)
        \label{eq:1}  \\
        h^\mu &=& \frac{1}{\sqrt N} \sum_j J_j \xi_j^\mu \,.
        \label{eq:2}
\end{eqnarray}
Here $g$ is the input--output relation of the perceptron, which is assumed
to be a monotonic non--decreasing function.  In (\ref{eq:1}) $\gamma$
denotes a gain parameter, and $h^\mu$ is the local field generated by  the
inputs $\{\xi_i^\mu\}$ as specified
in (\ref{eq:2}). The $J_j$ are couplings of an architecture of perceptron
type. We restrict our attention to general unbiased input patterns
specified  by $\langle \xi_i^\mu\rangle = 0$ and  $\langle \xi_i^\mu
\xi_j^\nu \rangle=\delta_{\mu,\nu} \delta_{i,j } C$. Since the effect of
$C$ in (\ref{eq:1}) can be absorbed in the gain
parameter we take $C=1$ in the sequel.

We explicitly allow a limited output precision in the mapping (\ref{eq:1}).
In other words the output that results when the input layer is in the
state $\{\xi_i^\mu\}$ is accepted if
\begin{equation}
        g(\gamma(h^\mu )) \in I_{\rm out}(\zeta^\mu,\epsilon)
                \equiv[\zeta^\mu-\epsilon,\zeta^\mu+\epsilon],
        \quad\mu=1,\dots,p\, ,
        \label{eq:3}
\end{equation}
where $\epsilon$ denotes the allowed output--error tolerance.

The strategy of the canonical approach is to require the graded--perceptron
network to go through a learning stage in the space of couplings in order
to find for the absolute minima of a given cost function
$E\left(\{h^{\mu}\},\{\zeta^{\mu}\}\right)$ precisely
networks with the properties (\ref{eq:3}). This cost function is assumed
to be a sum of local terms for each pattern $\mu$
\begin{equation}
        E\left(\{h^{\mu}\},\{\zeta^{\mu}\}\right)
                =\sum_{\mu} V\left(h^{\mu},\zeta^{\mu} \right)\,.
        \label{eq:4}
\end{equation}
The different cost functions that will be studied here can be put into the
form
\begin{equation}
        V\left(h^{\mu},\zeta^{\mu} \right)
                =W_s\left(\zeta^{\mu}-\epsilon-g\left(\gamma h^{\mu}\right)
                    \right)
                +W_s\left(g\left(\gamma h^{\mu}\right) - \zeta^{\mu} -
                 \epsilon\right) \,,
                 \label{eq:5}
\end{equation}
where
\begin{equation}
        W_s(x)=x^s\theta\left(x\right)\,,
\label{eq:6}
\end{equation}
and $\theta (x)$ is the Heaviside step function. For $s=0$ we get the GD
cost function,
that simply counts the number of the errors, irrespective of their size.
Moreover we consider a linear cost function ($s=1$), where the errors are
weighted proportionally to their magnitudes and a quadratic cost function
($s=2$) where the errors are weighted proportional to the square of their
magnitudes. The relevance of this choice becomes clear when
applying gradient descent dynamics
to eq.~(\ref{eq:4}) with the result
\begin{eqnarray}
     \Delta J_j&=&\frac{s\gamma\delta}{N} \sum_{\mu} \left(\xi_j^{\mu}
                -\frac{J_j h^{\mu}} {\sqrt N}\right)
                \left[W_{s-1}\left(\zeta^{\mu}-\epsilon- g\left(\gamma
            h^{\mu}\right)\right)\right.\nonumber\\
                & &\left.\quad +W_{s-1}\left(g\left(\gamma h^{\mu}\right)
      -\zeta^{\mu}-\epsilon\right)\right] g'\left(\gamma h^{\mu}\right)\,.
        \label{eq:7}
\end{eqnarray}
Taking $s=1$ respectively $s=2$ in this expression, we find back the
perceptron learning algorithm respectively the adaline learning algorithm
with step size $\delta$ for the graded perceptron.
The GD cost function does not correspond to any learning algorithm.

\section*{3. Replica theory}
\pin
The physical properties of the graded--response perceptron network defined
above are derived by investigating the canonical ensemble generated by the
free energy
\begin{equation}
        f(\beta)= -\lim_{N\rightarrow\infty} \frac{1}{N\beta}\ln Z\,,
        \label{eq:8}
\end{equation}
where $Z$ is the partition function
\begin{equation}
        Z=\int\prod_j{\rm d}J_j\prod_j\delta\left(\sum_j J_j^2 - N \right)
            \exp\left[-\beta E\left(\{h^{\mu}\},\{\zeta^{\mu}\}\right)
            \right]\,.
            \label{eq:9}
\end{equation}
In (\ref{eq:9}) the mean spherical constraint $\sum_i J_i^2=N$ is adopted
to fix a scale for the gain parameter $\gamma $ of the input--output
relation. We are interested in the limit $\beta \rightarrow \infty$ in
which the free energy gives information about
the fraction of patterns that are stored incorrectly. In the usual way the
free energy is assumed to be self--averaging with respect to the inputs
$\{\xi^{\mu}\}$ and the outputs $\{\zeta^{\mu}\}$. This average, denoted
by $\langle f(x) \rangle_{\{\xi^{\mu}\},\{\zeta^{\mu}\}} \equiv \langle f
\rangle$, can be performed by applying the replica trick.
The standard order parameter that appears in such a replica calculation is
the overlap between two distinct replicas in coupling space
\begin{equation}
        q_{\lambda \lambda'} \equiv {1 \over N} \sum_{i=1}^N J_i^\lambda
                J_i^{\lambda'} \quad \lambda < \lambda', \quad \lambda,
                \lambda'=1,\dots,n\,.
        \label{eq:10}
\end{equation}
In the sequel we consider both the replica symmetry (RS) analysis and the
one--step breaking effects (RSB1). We also suppress the index $\mu$.

In the RS analysis we assume that
\begin{equation}
        q_{\lambda \lambda'} = q\,, \quad \lambda < \lambda'\,.
        \label{eq:11}
\end{equation}
The optimal capacity properties of the system are obtained in the limit
$\beta \rightarrow \infty$, $q\rightarrow 1$, with $\beta (1-q)=x$ taking
a finite value. In this limit, a standard calculation analogous to the
binary perceptron problem \cite{GD,GG} leads to the averaged free--energy
\begin{equation}
        \langle f \rangle ={\rm extr}_{x} \left\{-\frac{1}{2x}
                +\alpha \left\langle\int {\rm D} t \min_{h}
                \left[F_{RS}\left(h,\zeta,x,t\right)\right]
               \right\rangle_{\{\zeta\}} \right\}\,,
               \label{eq:12}
\end{equation}
with
\begin{equation}
        F_{RS}(h,\zeta,x,t) = V(h,\zeta)+\frac{(h-t)^2}{2x}\,,
        \label{eq:13}
\end{equation}
and where ${\rm D}t=({\rm d}t/\sqrt{2\pi})\exp(-t^2/2)$, $\alpha =p/N$
denotes the storage capacity and $\langle\dots\rangle_{\{\zeta\}}$
indicates the average over the distribution of the output patterns.

Let us denote by $h_0(\zeta,x,t)$ the value of $h$ that
minimizes $F(h,\zeta,x,t)$. For a determined storage capacity  $\alpha $
the variable $x$
is given by the saddle point equation  $\partial\langle f\rangle/\partial
x=0$, that can be rewritten in the form
\begin{equation}
        \alpha^{-1}_{\rm RS} = \left\langle \int {\rm D} t
            \left(h_0(\zeta,x,t) - t\right)^2 \right\rangle_{\{\zeta\}}\,.
        \label{eq:14}
\end{equation}

We immediately remark that these results are not always stable against
RSB. Following standard considerations \cite{GD,AT,Bouten} the stability
condition reads
\begin{equation}
        \alpha_{RS}\left\langle\int {\rm D}t \left[\frac{\rm d}{{\rm d}t}
                \left[h_0(\zeta,x,t) - t\right]\right]^2\right
               \rangle_{\{\zeta\}} <1 \,.
               \label{eq:15}
\end{equation}

For the exact mapping task where $\epsilon =0$ the result found in
\cite{BKvM} for the critical storage capacity corresponding to the
GD cost function is retrieved when we take the limit
$x\rightarrow\infty$ in (\ref{eq:14})
\begin{equation}
        \alpha_{c}^{-1} = 1+\langle h_{\zeta}^2 \rangle_{\{\zeta\}}\,,
        \label{eq:16}
\end{equation}
with
\begin{equation}
        h_{\zeta} = \frac{1}{\gamma}g^{-1}\left(\zeta\right)\,.
        \label{eq:17}
\end{equation}

Similar to binary networks \cite{GG,MEZ}, $\alpha_c$ is the same for all
cost functions. Clearly, for $\alpha>\alpha_c$ errors will be introduced
that depend both in quantity and in size on the specific cost function
used. An interesting expression to look
at in this respect is the distribution of local fields since it provides
more information on the deviation of the errors from the correct output
$\zeta $. For a given desired output $\zeta$, it is defined as
\begin{equation}
        \rho(h|\zeta)=\left\langle\delta\left(h-\frac{1}{\sqrt{N}}
         \sum_{j=1}^N J_j\xi_j\right)\right\rangle_{\{J\},\{\xi\}}\,,
        \label{eq:18}
\end{equation}
where the thermal average over $J$ is taken subject to the mean spherical
constraint introduced before. Following Kepler and Abbott \cite{KA},
we find for the graded perceptron
\begin{equation}
     \rho_{RS}(h|\zeta)=\int{\rm D}t \,\,
                   \delta\left(h-h_0(\zeta,x,t)\right)\,.
         \label{eq:19}
\end{equation}
An overall measure of the network performance is given by the average
output error
\begin{equation}
        {\cal E}=\left\langle {\cal E}(\zeta) \right\rangle_{\{\zeta\}}\,,
        \label{eq:20}
\end{equation}
where the $\zeta$--dependent output error ${\cal E}(\zeta)$ is given by
\begin{equation}
        {\cal E}(\zeta)=\int {\rm d}h \,\,\rho_{RS} (h|\zeta)\,\,
                \left[W_1\left(\zeta - \epsilon -  g\left(\gamma
           h\right)\right) +W_1\left(g\left(\gamma h\right) - \zeta
                -\epsilon\right)\right]\,.
        \label{eq:20a}
\end{equation}

From the results in the literature on the binary perceptron problem
\cite{ET,MEZ,WS} and from our former studies on the graded perceptron
system \cite{BKvM,BE} we expect RSB effects. So we want to improve the  RS
results by applying the first step of Parisi's
RSB scheme \cite{MPV}. We, therefore, introduce the following order
parameters
\begin{eqnarray}
        q_{\lambda \lambda'} = q_{\beta_1 \beta_2 }^{\alpha_1 \alpha_2 }
                = \left\{ \begin{array}{ll}
                q_1 & \mbox{if} \quad \alpha_1= \beta_1 \\
              q_0 & \mbox{if} \quad \alpha_1 \neq \beta_1 \end{array}\right.
        \label{eq:21}
\end{eqnarray}
where $\alpha_1, \beta_1=1,\dots,n/m; \alpha_2, \beta_2=1,\dots,m $ and  $1
\leq m \leq n$. We remark that in the limit $n \rightarrow 0$,  $0 \leq m
\leq 1$.

Similar to \cite{MEZ} we find after a standard but tedious calculation
that in the limit $q_1 \rightarrow 1^{-}, m \rightarrow 0$ and  $0 \leq q_0
\leq q_1$ with $m/(1-q_1) =M$ a finite value and  $x =
\beta\left(1-q_1\right)$, the free energy averaged with respect to the
inputs $\{\xi\}$ and the outputs $\{\zeta\}$ can be written as
\begin{eqnarray}
        \langle f \rangle
        &=& \lim_{\beta \rightarrow \infty} \,\, \max_{x,q_0,M}
       \left\{-\frac{1}{2Mx} \ln [1+M (1-q_0)]\right. \nonumber\\
        &-& \left.\frac{q_0}{2x\left[1 + M (1-q_0)\right]}
             - \frac{\alpha}{Mx} \left\langle \int {\rm D}t_0  \ln
        \Psi(\zeta,x,q_0,M,t_0) \right\rangle_{\{\zeta \}}\right\}
        \label{eq:22}
\end{eqnarray}
with
\begin{equation}
        \Psi(\zeta ,x,q_0,M,t_0)=\int {\rm D}t_1\exp\left\{-Mx \min_h
                F_{RSB1}(h,\zeta,x,q_0,t_0,t_1) \right\}
        \label{eq:23}
\end{equation}
and
\begin{equation}
        F_{RSB1}(h,\zeta ,x,q_0,t_0,t_1) = V(h,\zeta) +
        \frac{1}{2x}
            \left(h - t_0\sqrt{q_0} -t_1\sqrt{1-q_0}\right)^2 \,.
        \label{eq:24}
\end{equation}
For a chosen storage capacity $\alpha$, the variables $x$, $q_0$ and
$M$ are given by the saddle point equations $\partial\langle
f\rangle/\partial x=0$, $\partial\langle f\rangle/\partial q_0=0$  and
$\partial\langle f\rangle/\partial M=0$.

The first step RSB distribution for the local fields corresponding  to
pattern $\zeta $ becomes
\begin{equation}
    \rho_{RSB1}(h,\zeta)=\int{\rm D}t_0 \int{\rm D}t_1
        \frac{\exp \left[-Mx F_{RSB1}
                \left(h_0,\zeta,x,q_0,t_0,t_1\right)\right]
          \delta\left(h-h_0\right)}{\Psi(\zeta ,x,q_0,M,t_0)} \,,
          \label{eq:25}
\end{equation}
where $h_0=h_0(\zeta,x,q_0,t_0,t_1)$ is the value of $h$ that minimizes
$F_{RSB1} \left(h,\zeta,x,q_0,t_0,t_1\right)$.
The average output error in RSB1 approximation is obtained by replacing the
expression (\ref{eq:19}) by (\ref{eq:25}) in (\ref{eq:20a}).

\section*{4. Results for specific cost functions}
\pin
The theory outlined in the last Section has been applied to the specific
cost functions defined in (\ref{eq:5})-(\ref{eq:6}).  For the input--output
relation $g$ we have used both the hyperbolic tangent and the
piecewise--linear function
\begin{eqnarray}
        g(x)=\left \{\begin{array}{ll}
                x & \mbox{for} \quad  |x| < 1 \\
                {\rm sign}(x) & \mbox{elsewhere}
                \end{array} \right.\,.
        \label{eq:26}
\end{eqnarray}
A priori, our aim is not to compare the macroscopic properties of
graded--perceptrons for the two different input--output relations since, in
general, they are qualitatively the same. In fact, the results obtained
here are complementary. For the hyperbolic tangent input--output relation
the RS solution is found to be stable over an important
range of values for the parameters $\alpha$ and $\gamma$ while
in the case of the piecewise--linear input--output relation the RS solution
is always unstable. However, from a more technical point of view in the case
of the hyperbolic tangent function, the
mimimization of $F_{RS}$ in the corresponding averaged RS free--energy with
respect to the local field $h$ (recall eqs.~(\ref{eq:12}) and
(\ref{eq:13})) only leads to an equation defining $t$ as a
function of the minimizing value $h_0$ (see the Appendix). This equation
needs
to be inverted but depending on the values of $\gamma^2 x$ and $\zeta$ the
inverse function may be multiple--valued and hence a (sometimes very
tedious) Maxwell construction is required in order to make it
single--valued. Consequently, only the RS solution is studied in detail in
this case. On the contrary, the piecewise--linear input--output relation
permits an explicit calculation of the minimizing values
$h_0(\zeta,x,t)$ and $h_0(\zeta,x,q_0,t_0,t_1)$ of the functions $F_{RS}$
and $F_{RSB1}$ in the corresponding averaged free--energies. This, in turn,
simplifies drastically the calculations and both the RS and RSB1 solutions
are completely worked out in this case.

At this point, we remark already that the Maxwell construction in the
hyperbolic tangent case gives a discontinuity in $h_0(t)$ having an effect
on the stability of the RS solution. Similarly, due to the fact that the
piecewise--linear input--output relation is not everywhere differentiable a
gap structure in the distribution of the local fields emerges
signalling the instability of the RS solution \cite{Bouten}.
The effects of RSB for the cost functions
(\ref{eq:5})-(\ref{eq:6}) are found to be important.

In the sequel we present the results of our calculations both for the
hyperbolic tangent and the piecewise--linear input--output relations. In
order not to interrupt the line of reasoning we refer all technical details
of the calculations to the Appendix.

\subsection*{4.1. Hyperbolic tangent input-output relation}
\pin
In this part we compare the performance of the three cost functions defined
in (\ref{eq:5})-(\ref{eq:6}) by studying their average output error, $\cal
E$ (recall eq.~(\ref{eq:20})). Our strategy is to consider a linear $(s=1)$
and quadratic $(s=2)$ ``entirely soft'' cost function versus a
``completely rigid'' one $(s=0)$. Soft means that we do not fix the
output-error tolerance
$\epsilon$, since some outputs might be far away from the correct output
$\zeta$. Entirely soft indicates that we work without tolerance at all
by putting $\epsilon=0$. For the
completely rigid cost function, $\epsilon$ was determined in function of
the loading capacity $\alpha$, by solving (for $\epsilon$) the optimal
capacity for the graded perceptron in the microcanonical approach (recall
eq.~(9) of ref.~\cite{BKvM}).

The results are presented in Figs.~1 and 2. First, we show in Figs.~1a--c
the
loading capacity $\alpha$ as a function of the gain parameter $\gamma$ for
a constant average output error ${\cal E}=0, 0.1$ and $0.2$ in the case of
the three cost functions. For the rigid cost function, we plot an
additional curve for ${\cal E}=0.4$ to indicate that
the capacity has a maximum for finite $\gamma$, although only for higher
values of $\cal E$. In the case of both the linear and
the quadratic cost functions no maximum
is found for a finite gain parameter.
Furthermore the de Almeida--Thouless (AT) line, $\alpha_{AT}$,
is given, indicating that the region of RS breaking (at the right of the
line) is important. The rigid cost function has the worst performance for
all values of $\gamma$.
For both the linear and the quadratic cost function a monotonically
increasing (but bounded) capacity $\alpha$ results. For all values of 
$\gamma$,
the linear cost function has the best performance.

This behaviour of the graded perceptron network can be understood in terms
of the ``strategy'' used by a specific cost function to arrange the local
fields when learning the patterns. The rigid cost function puts all local
fields in a connected interval, thereby minimizing its width. It does not
try to optimize the learning {\em inside} the interval in order to decrease
the average output error. However, the linear and quadratic soft cost
functions do optimize their performance by penalizing the errors linearly
respectively quadratically with their size. They try to arrange the local
fields in a close region around the value $h_{\zeta}$ resulting in the
correct output $\zeta$ under the action of the input-output relation.
In both cases the resulting distribution of the local fields shows a
sharp peak (a $\delta$--peak in the linear case) at $h_{\zeta}$, and
decreasing tails. A gap in between can occur. The tails of the quadratic
cost function decrease faster than those of the linear cost function. We
will present figures below for the case of the piecewise--linear
input--output relation where a similar behaviour has been found.

Finally it is very interesting to discuss in more detail the ``gap''
structure of the local fields, revealed by the line $\alpha_g$ in
Figs.~2a--b. For the rigid cost function, no gaps are present, since the
output tolerance $\epsilon$ is chosen such that {\em all} fields are inside
a connected interval. For the linear and  quadratic cost functions
Figs.~ 2a--b present the relevant results in the $(\alpha-\gamma)$--plane.
A gap is present in the region between the lines $\alpha_c$ (for ${\cal
E}=0$) and $\alpha_g$. For $\alpha<\alpha_c$, the perceptron is not
saturated, i.e., $q<1$ and the present calculations do not cover this
region. We notice that for small $\alpha $ the gap line lies very close to
the AT--line. A similar behaviour has been noticed in binary networks
trained with noisy patterns \cite{WoSh93}. For growing $\alpha \geq
\alpha_c$, the width of the gaps
decreases from an infinite value at $\alpha_c$ to become zero as $\alpha$
approaches $\alpha_g$. In the region between $\alpha_g$ and $\alpha_{AT}$
there are no gaps, but the RS solution remains unstable.

Concerning the stability of our results with respect to RS breaking,
we see that for the rigid cost function the curves for the capacity as a
function of the gain parameter at constant average output error are
``stable'' starting from $\gamma=0$ up to the point where the curves reach
their maximum (in agreement with the results of \cite{BKvM} for constant
output-error tolerance). For the linear cost function, the RS curves are
stable for small $\gamma$ and not so small $\cal E$. However, for the
quadratic cost function all the curves for the hyperbolic tangent
input-output relation are RS unstable.

The origin of instability against RS breaking fluctuations is
relatively easy to understand in the region where gaps in the local
field distribution are present \cite{GG}, \cite{Bouten}, \cite{WoSh}.
One can argue that it is not possible to pass continuously from one
replica of the system where a specific pattern is learned in one ``band''
of the local fields, to another replica where that pattern is learned in
another band. The corresponding solutions are disconnected in
the space of replicas, and the overlap between pairs of replicas cannot
be the same for all pairs, contrary to the RS assumption.
In the region where there are no gaps this argument is no longer valid.
Here, one may argue that spreading the local fields over one single but
wide band can also disrupt the space of replicas. May be the notion of
critical band width is relevant here. This could be an interesting subject
for further study.

\subsection*{4.2. Piecewise linear input--output relation}
\pin
For the piecewise linear input--output relation we do consider a non--zero
output--error tolerance $\epsilon$, i.e., all the inputs whose
corresponding output lie inside the interval $[\zeta-\epsilon,
\zeta+\epsilon]$ do not contribute to the average output error. As outlined
before the numerical calculations are easier than those for the hyperbolic
tangent, and the study of the RSB1 solution in some detail becomes
feasible. Numerical results are presented for $\epsilon=0.5$.

Before passing to these results, it is worth mentioning that the
introduction of a fixed $\epsilon$ allows us to replace the study of
the $s=0$ cost function with a completely rigid constraint discussed
in Section~4.1, by a true GD cost--function ((\ref{eq:5})-(\ref{eq:6})
for $s=0$).

In fig.~3 we see both the RS and the RSB1 average output error
${\cal E}$ as a function of the loading capacity $\alpha$ for $\gamma=1$
for the three cost functions considered. As expected, ${\cal E}_{ RSB1} >
{\cal E}_{RS}$ for all $\alpha>\alpha_c$. In the present region of
the network parameters, the linear cost--function gives the best
performance. According to the ${\rm RSB}_1$ results, the least efficient is 
the
quadratic cost--function if $\alpha<0.48$, and the GD cost--function 
elsewhere.

Figure~4 shows $\alpha$ as a function of $\gamma$ at constant $\cal E$. For
each cost--function, the upper (lower) curve corresponds to the RS
(RSB1) result. For all $\gamma$ the highest capacity is given by
the linear cost function and for $\gamma < \pm 2.5$, the quadratic cost
function gives the lowest capacity.

The reason that the performance of the $s=2$ quadratic cost function
is worse here is based on the fact that with a non--zero
$\epsilon$, the average output error decreases. The curves for the
hyperbolic tangent input--output relation are all for ${\cal E}\leq 0.4$,
while for the piecewise linear input--output relation
we have studied the case ${\cal E}=0.05$. In
the latter, we are closer to the critical capacity. From these calculations
one might conclude that the relative performance of the different
cost--functions depends also on the amount of errors. In other words, it
matters how far one is beyond the critical capacity and the quadratic
cost--function performs better in the high--$\alpha$ regime.

In order to discuss in more detail the effects of RSB, we have studied the
distribution of the local fields for the three cost--functions. As an
illustrative example, we present in fig.~5a--c the RS and RSB1
distributions for the specific parameters $\alpha=3$, $\gamma=1$,
$\epsilon=0.5$ and $\zeta=0.6$.
In general, the discussion above concerning the RS field distribution
for the hyperbolic tangent input--output relation remains valid. For the
RSB1 distribution, the following has to be remarked. In the case of the GD
and the linear cost--function the coefficients of the $\delta $-part in
the RSB1 local field distributions become smaller. To give an idea about
this change we mention that, e.g., for the GD cost function (recall
equations (\ref{eq:rhors}) and (\ref{eq:rhorsb})) these coefficients are
$0.479$ at $h=1$ for the RS
solution versus $0.306$ for the RSB solution. Similarly for the quadratic
cost function the maximum in the distribution decreases. Furthermore for
the three cost--functions, the continum part of the distribution is more
populated for the RSB1 than for the RS solution, and the width of
the gaps are smaller. Finally, RSB--effects in the local field
distibution are less pronounced for the quadratic cost--function.

\section*{5. Concluding remarks}
\pin
In this paper we have studied the canonical ensemble approach to the
optimal capacity of graded--response perceptrons with a hyperbolic tangent
and a piecewise--linear input--output relation for three different cost
functions: the Gardner-Derrida cost function that simply counts the number
of errors irrespective of their sizes, the linear cost function where
the errors are weighted proportionally to their magnitudes and the
quadratic
cost function where the errors are weighted proportionally to the square of
their magnitudes. Results have been obtained for the storage capacity as a
function of the gain parameter, for the distribution of the local fields
and for the average total output error above critical capacity in both RS
and RSB1 approximation.

The transition from RS to RSB occurs at the critical storage.
RSB1 effects are important, especially for the distributions of the
local fields. In agreement with standard results
it is seen that whenever the distribution displays a gap the RS saddle
point is certainly unstable. But in all cases considered here the
instability stays in regions of the network parameters where no gap occurs
(but, of course, the replicon eigenvalue is still positive).
For small loading the gap line lies very close to the AT--line.
The width of the gap itself decreases in RSB1.
Already for a small average total output error (and an output tolerance
$0.5$) the capacity is overestimated
in RS by typically about $10 \%$. In general, RSB1 effects are the smallest
for the quadratic cost function.

\section*{Acknowledgments}
\pin
This work has been supported in part by the Research Fund of the K.U.Leuven
(grant OT/94/9). We are indebted to R.~K\"uhn and J.~van Mourik for
stimulating discussions. DB thanks the Fund for Scientific
Research-Flanders (Belgium) for financial support. RE is supported in part
by the CNPq (Conselho Nacional de Desenvolvimento Cient\'{\i}fico e
Tecnol\'ogico), Brazil.

\renewcommand{\theequation}{\thesection.\arabic{equation}}
\appendix
\setcounter{equation}{0}
\section{Theory for specific cost functions}
\pin
In this appendix we apply the general theory discussed in Section~3 to the
specific cost functions (\ref{eq:5})-(\ref{eq:6}). In particular we study
the RS solutions for the hyperbolic tangent input--output relation and both
the RS and RSB1 solutions for the piecewise--linear input--output
relation defined in (\ref{eq:26}).

\subsection{GD cost function}
\pin
In the case of the GD cost function with output tolerance $\epsilon$ the
results presented here are valid, of course, for both input--output
relations considered, by taking in the end the relevant expression for
$g^{-1}(\zeta-\epsilon)$ .
In the case of the RS treatment, the minimum in $h$ of Eq. (\ref{eq:13}) is
given by
\begin{eqnarray}
    \begin{array}{lll}
            h_0=t\,, & \quad  F_{RS}(h_0,\zeta,x,t)=1 & \mbox{for}
               \quad -\infty<t<l-\sqrt{2x} \\
            h_0=l\,,&\quad  F_{RS}(h_0,\zeta,x,t)=\frac{(l-t)^2}{2x} &
                    \mbox{for} \quad  l-\sqrt{2x}<t<l \\
            h_0=t\,,&\quad  F_{RS}(h_0,\zeta,x,t)=0 & \mbox{for}  \quad
                    l<t<u \\ h_0=u\,,
            &\quad  F_{RS}(h_0,\zeta,x,t)=\frac{(u-t)^2}{2x} &
                    \mbox{for} \quad  u<t<u+\sqrt{2x} \\
            h_0=t\,,&\quad  F_{RS}(h_0,\zeta,x,t)=1 & \mbox{for}  \quad
                        u+\sqrt{2x}<t<\infty
        \end{array}
        \label{eq:27}
\end{eqnarray}
where
\begin{eqnarray}
        l = \left\{ \begin{array}{ll}
        \frac{1}{\gamma}g^{-1}(\zeta-\epsilon) & \mbox{if}
        \quad \zeta-\epsilon > -1 \\
        -\infty  & \mbox{elsewhere} \end{array}\right.
        \label{eq:28}
\end{eqnarray}
and
\begin{eqnarray}
        u = \left\{ \begin{array}{ll}
        \frac{1}{\gamma}g^{-1}(\zeta+\epsilon) & \mbox{if}
        \quad \zeta+\epsilon < 1 \\
        \infty  & \mbox{elsewhere}\,. \end{array}\right.
        \label{eq:29}
\end{eqnarray}
>From (\ref{eq:14}) and (\ref{eq:27}), we obtain the saddle--point equation
\begin{equation}
        \alpha_{RS}^{-1} = \left\langle\int_{l-\sqrt{2x}}^l {\rm D}t (t-l)^2
        + \int_u^{u+\sqrt{2x}} {\rm D}t (t-u)^2\right\rangle_{\zeta}\,.
        \label{eq:30}
\end{equation}
Combining (\ref{eq:19}) and (\ref{eq:27}), the distribution of local fields
becomes
\begin{eqnarray}
        \rho(h,\zeta)
        &=& \frac{{\rm e}^{-\frac{h^2}{2}}}{\sqrt{2 \pi}}
                \left[\theta(l-\sqrt{2x}-h) + \theta(h-l)-\theta(h-u)+
                \theta(h-u-\sqrt{2x})\right] \nonumber\\
        &+& \delta(h-l)\int_{l-\sqrt{2x}}^l {\rm D}t
                + \delta(h-u)\int_u^{u+\sqrt{2x}} {\rm D}t\,.
        \label{eq:rhors}
\end{eqnarray}
The RS output error is obtained from (\ref{eq:20a}) and (\ref{eq:rhors}):
\begin{eqnarray}
        {\cal E}(\zeta) &=& \gamma \left[\left(l+\frac{1}{\gamma}\right)
                \int_{-\infty}^{h_1}{\rm D}h + \int_{h_1}^{l-
                \sqrt{2x}}{\rm D}h (l-h)\right.\nonumber\\
                &+& \left.\int_{u+\sqrt{2x}}^{h_2}{\rm D}h (h-u)
                +\left(\frac{1}{\gamma}-u \right)
                \int_{h_2}^{-\infty}{\rm D}h\right]\,,
        \label{eq:32}
\end{eqnarray}
where
\begin{equation}
        h_1={\rm min}\left(l-\sqrt{2x},-\frac{1}{\gamma}\right)
        \label{eq:33}
\end{equation}
and
\begin{equation}
        h_2={\rm max}\left(u+\sqrt{2x},\frac{1}{\gamma}\right)\,.
        \label{eq:34}
\end{equation}

For the RSB1 solution we get $h_0(\zeta,x,q_0,t_0,t_1)$ and
$F_{RSB1}(h_0,\zeta,x,q_0,t_0,t_1)$ from $h_0(\zeta,x,t)$ and
$F_{RS}(h_0,\zeta,x,t)$, respectively, by substituting $t$ by
$t_0\sqrt{q_0}+t_1\sqrt{1-q_0}$ in (\ref{eq:27}).
The function $\Psi(\zeta,x,q_0,M,t_0)$ in (\ref{eq:23}) becomes
\begin{eqnarray}
& &\Psi(\zeta,x,q_0,M,t_0)={\rm e}^{-Mx}\int_{-\infty}^
              {\Omega(l-\sqrt{2x},q_0,t_0)}{\rm D}t_1\nonumber\\
      & &\quad+\int_{\Omega(l-\sqrt{2x},q_0,t_0)}^{\Omega(l,q_0,t_0)}{\rm 
D}t_1
                \,\Phi(l,M,q_0,t_0,t_1)
        +\int_{\Omega(l,q_0,t_0)}^{\Omega(u,q_0,t_0)}{\rm D}t_1\\
      &+&\int_{\Omega(u,q_0,t_0)}^{\Omega(u+\sqrt{2x},q_0,t_0)}{\rm D}t_1
      \,\Phi(u,M,q_0,t_0,t_1)+{\rm e}^{-Mx}\int_{\Omega(u+\sqrt{2x},q_0,t_0)}
                ^{\infty}{\rm D}t_1\nonumber\,,
        \label{eq:35}
\end{eqnarray}
where
\begin{equation}
\Omega(\omega,q_0,t_0)=\frac{\omega-t_0\sqrt{q_0}}{\sqrt{1-q_0}}
\label{eq:35a}
\end{equation}
and
\begin{equation}
\Phi(\omega,M,q_0,t_0,t_1)=\exp\left\{-\frac{1}{2}M(1-q_0)
        \left[\Omega(\omega,q_0,t_0)-t_1\right]^2\right\}\,.
\label{eq:35b}
\end{equation}
The averaged free--energy is obtained by plugging this expression into
(\ref{eq:22}).
Expression (\ref{eq:25}) then leads to the $\zeta$--dependent distribution
of local fields
\begin{eqnarray}
 \rho(h,\zeta)&=&\int\frac{{\rm D}t_0}{\Psi(\zeta,x,q_0,M,t_0)}
  \Biggl\{\frac{{\rm exp}\left[-\frac{1}{2}\Omega^2(h,q_0,t_0)\right]}
        {\sqrt{2\pi(1-q_0)}}\left[{\rm e}^{-Mx}\theta\left(l-\sqrt{2x}-
h\right)
   \right. \nonumber\\
        &+&\left.\left[\theta\left(h-l\right)-\theta\left(h-u\right)
        \right]+{\rm e}^{-Mx}\theta\left(h-u-\sqrt{2x}\right)\right]
        \nonumber\\
    &+&\delta(h-l)\int_{\Omega(l-\sqrt{2x},q_0,t_0)}^{\Omega(l,q_0,t_0)}
        {\rm D}t_1\,\Phi(l,M,q_0,t_0,t_1)\nonumber\\
    &+&\delta(h-u)\int_{\Omega(u,q_0,t_0)}^{\Omega(u+\sqrt{2x},q_0,t_0)}
        {\rm D}t_1\,\Phi(u,M,q_0,t_0,t_1)\Biggr\}\,.
        \label{eq:rhorsb}
\end{eqnarray}
Finally, the $\zeta$--dependent RSB1 output error is given by combining
(\ref{eq:25}) and (\ref{eq:rhorsb}):
\begin{eqnarray}
     {\cal E}(\zeta)&=&\int\frac{{\rm D}t_0}{\Psi(\zeta,x,q_0,M,t_0)}
         \frac{\gamma\,{\rm e}^{-Mx}}{\sqrt{2\pi(1-q_0)}}
          \left\{\left(l+\frac{1}{\gamma}\right)
                  \int_{-\infty}^{h_1}{\rm d}h
  +\int_{h_1}^{l-\sqrt{2x}}{\rm d}h\,(l-h)\right.\nonumber\\
        &+&\left.\int_{u+\sqrt{2x}}^{h_2}{\rm d}h\,(h-u)
  +\left(\frac{1}{\gamma}-u\right)\int_{h_2}^{\infty}{\rm d}h\right\}
   \exp\left[-\frac{1}{2}\Omega^2(h,q_0,t_0)\right]\, ,
        \label{eq:37}
\end{eqnarray}
with $h_1$ and $h_2$ defined in (\ref{eq:33}) and (\ref{eq:34})
respectively.

\subsection{Linear cost function}
\pin
Let us start by considering the RS approximation first.
For the piecewise linear input--output relation the minimization in $h$ of
Eq.~(\ref{eq:13}) can be done explicitly leading to the following result
\begin{eqnarray}
    \begin{array}{lll}
     h_0=t\,, & \quad  F_{RS}(h_0,\zeta,x,t)=
         \gamma\left(l+\frac{1}
               {\gamma}\right) & \mbox{for} \quad -\infty<t<h_1 \\
     h_0=\gamma x+t\,, & \quad  F_{RS}(h_0,\zeta,x,t)=\gamma\left(l
                  -\frac{\gamma x}{2}-t\right) & \mbox{for} \quad
                        h_1<t<h_2 \\
     h_0=l\,,&\quad  F_{RS}(h_0,\zeta,x,t)=\frac{(l-t)^2}{2x} &
                  \mbox{for} \quad h_2<t<l \\
     h_0=t\,,&\quad  F_{RS}(h_0,\zeta,x,t)=0 & \mbox{for} \quad  l<t<u \\
     h_0=u\,,&\quad  F_{RS}(h_0,\zeta,x,t)=\frac{(u-t)^2}{2x} &
                        \mbox{for} \quad  u<t<h_3 \\
     h_0=-\gamma x+t\,,&\quad  F_{RS}(h_0,\zeta,x,t)=\gamma\left(-u
                        -\frac{\gamma x}{2}+t\right) & \mbox{for}
                        \quad u<t<h_4\\
     h_0=t\,, & \quad  F_{RS}(h_0,\zeta,x,t)=\gamma\left(\frac{1}{\gamma}
                        -u\right) & \mbox{for} \quad h_4<t<\infty \, ,
        \end{array}
        \label{eq:38}
\end{eqnarray}
where $l$ and $u$ are again given by the formula (\ref{eq:28}) and
(\ref{eq:29}). The variables $h_1$, $h_2$, $h_3$ and $h_4$ are defined as
follows:
\begin{eqnarray}
        h_1=\left \{\begin{array}{ll}
                l-\sqrt{2\gamma x\left(l+\frac{1}{\gamma}\right)}
            &\mbox{if}\quad l< -\frac{1}{\gamma}+\frac{\gamma x}{2}\\
             -\frac{1}{\gamma}-\frac{\gamma x}{2} & \mbox{elsewhere}
                \end{array} \right.\, ,
        \label{eq:39}
\end{eqnarray}
\begin{eqnarray}
     h_2=\left \{\begin{array}{ll}
                l-\sqrt{2\gamma x\left(l+\frac{1}{\gamma}\right)}
               &\mbox{if}\quad l< -\frac{1}{\gamma}+\frac{\gamma x}{2}\\
                l-\gamma x & \mbox{elsewhere}
                \end{array} \right.\, ,
        \label{eq:40}
\end{eqnarray}
\begin{eqnarray}
        h_3=\left \{\begin{array}{ll}
          u+\sqrt{2\gamma x\left(\frac{1}{\gamma}-u\right)}
               &\mbox{if}\quad u > \frac{1}{\gamma}-\frac{\gamma x}{2}\\
                u+\gamma x & \mbox{elsewhere}
                \end{array} \right.\, ,
        \label{eq:41}
\end{eqnarray}
\begin{eqnarray}
        h_4=\left \{\begin{array}{ll}
                u+\sqrt{2\gamma x\left(\frac{1}{\gamma}-u\right)}
               &\mbox{if}\quad u> \frac{1}{\gamma}-\frac{\gamma x}{2}\\
                \frac{1}{\gamma}+\frac{\gamma x}{2} & \mbox{elsewhere}
                \end{array} \right.\, .
        \label{eq:42}
\end{eqnarray}
The RS saddle--point equation is obtained from (\ref{eq:14}) and
(\ref{eq:38}):
\begin{equation}
     \alpha_{RS}^{-1} = \left\langle
     \gamma^2 x^2\int_{h_1}^{h_2} {\rm D}t+\int_{h_2}^l {\rm D}t (t-l)^2
     +\int_u^{h_3} {\rm D}t (t-u)^2+\gamma^2 x^2\int_{h_3}^{h_4} {\rm D}t
               \right\rangle_{\zeta}\,.
        \label{eq:43}
\end{equation}
Using (\ref{eq:19}) and (\ref{eq:38}), the RS $\zeta$--dependent
distribution of local fields becomes
\begin{eqnarray}
  \rho(h,\zeta)
   &=& \frac{\exp\left[{-\frac{h^2}{2}}\right]}{\sqrt{2 \pi}}
       \left[\theta(h_1-h) +\theta(h-l)-\theta(h-u)+\theta(h-h_4)\right]
            \nonumber\\
   &+& \frac{\exp\left[{-\frac{(h-\gamma x)^2}{2}}\right]}{\sqrt{2\pi}}
      \left[\theta(h-h'_2)-\theta(h-l)\right]+\delta(h-l)\int_{h_2}^l
           {\rm D}t \nonumber\\
   &+& \delta(h-u)\int_u^{h_3} {\rm D}t
      +\frac{\exp\left[{-\frac{(h+ \gamma x)^2}{2}}\right]}
      {\sqrt{2\pi}}\left[\theta(h-u)-\theta(h-h'_3) \right]\, ,
        \label{eq:44}
\end{eqnarray}
where
\begin{eqnarray}
    h'_2=\left \{\begin{array}{ll}
          l &\mbox{if}\quad l< -\frac{1}{\gamma}+\frac{\gamma x}{2}\\
             -\frac{1}{\gamma}+\frac{\gamma x}{2} & \mbox{elsewhere}
               \end{array} \right.\,,
        \label{eq:45}
\end{eqnarray}
and
\begin{eqnarray}
    h'_3=\left \{\begin{array}{ll}
           u &\mbox{if}\quad u > \frac{1}{\gamma}-\frac{\gamma x}{2}\\
               \frac{1}{\gamma}-\frac{\gamma x}{2} & \mbox{elsewhere}
               \end{array} \right.\,.
        \label{eq:46}
\end{eqnarray}
The RS average output error is obtained from (\ref{eq:20a}) and
(\ref{eq:44}):
\begin{eqnarray}
    {\cal E}(\zeta) &=& \gamma \left[\left(l+\frac{1}{\gamma}\right)
           \int_{-\infty}^{h_1}\frac{{\rm d}h}{\sqrt{2\pi}}{\rm e}^
              {-\frac{h^2}{2}}+
              \int_{h'_2}^l{{\rm d}h}{\sqrt{2\pi}} \exp\left[
              {-\frac{(h-\gamma x)^2}{2}}\right](l-h)
                \right.\nonumber\\
    &+&\left.\int_u^{h'_3}\frac{{\rm d}h}{\sqrt{2\pi}} \exp\left[
      {-\frac{(h+\gamma x)^2}
         {2}}\right](h-u)+\left(\frac{1}{\gamma}-u\right)\int_{h_4}^
               {-\infty}\frac{{\rm d}h}
               {\sqrt{2\pi}}{\rm e}^{-\frac{h^2}{2}}\right]
        \nonumber\\ \label{eq:47}
\end{eqnarray}

For the hyperbolic tangent input--output relation, $h_{\zeta}$ given by
(\ref{eq:17}) is always a local minimum of (\ref{eq:13}). Other local
minima of (\ref{eq:13}) are defined as solutions of
\begin{equation}
     \left(\frac{\partial F(h,\zeta,x,t)}{\partial h}\right)_{h=h_0}=0
      \label{eq:h30}
\end{equation}
and they can no longer be determined analytically. The equation
(\ref{eq:h30}) defines $t$ as a function of $h_0$,
\begin{equation}
     t(h_0)=h_0+\gamma x g'(\gamma h_0){\rm sgn}\left[g(\gamma h_0)
      -h\right]  \,,
      \label{eq:h31}
\end{equation}
that needs to be inverted in order to find $h_0=h_0(t)$ (the prime denotes
the derivative with respect to $h$). Depending on the value of
$\gamma^2 x$, $t(h_0)$ is a monotonic function or not, and consequently
it is invertible or not. The onset of non--monotonicity is given by the
system of equations
\begin{eqnarray}
          \left \{
               \begin{array}{l}
                \frac{{\rm d}t}{{\rm d}h_0}=0\\
                \frac{{\rm d}^2 t}{{\rm d}h_0^2}=0 \,.
          \end{array}
           \right.
           \label{eq:h32}
\end{eqnarray}
If monotonicity holds, $h_0$ is a solution of (\ref{eq:h31}) for
$t<t_{\zeta}^{-}$ or $t>t_{\zeta}^{+}$, where $t_{\zeta}^{\pm}=h_{\zeta}\pm
\gamma x g'(\gamma h_{\zeta})$. If non-monotonicity holds, $h_0(t)$ has one
or two jumps at $t=t_1$ and/or $t=t_2$, whereby we assume that $t_1<t_2$.
The values of $t_1$ and $t_2$ are then determined using a
Maxwell construction in the function $t(h_0)$. The number of jumps depends
on the value of $\gamma^2 x$ and $\zeta$.

>From (\ref{eq:19}) and from the inversion of (\ref{eq:h31}), we obtain the
following expression for the distribution of the local fields:
\begin{equation}
  \rho(h|\zeta)=\frac{{\rm d}t}{{\rm d}h}
            \frac{\exp\left[{-\frac{t^2 (h)}{2}}\right]}{\sqrt{2\pi}}
             +\frac{1}{2}\left(\erf\left(\frac{t_2}{\sqrt{2}}\right)-
                \erf\left(\frac{t_1}{\sqrt{2}}\right)\right)
                \delta(h-h_{\zeta})      \,.
      \label{eq:h33}
\end{equation}
If monotonicity holds, then $t_1=t_{\zeta}^{-}$ and $t_2=t_{\zeta}^{+}$.
Due to the fact that $h_{\zeta}$ is a global minimum of (\ref{eq:13}) in
the interval $t_1<t<t_2$, $t(h_0)$ always displays a jump in
$h=h_{\zeta}$. This jump gives rise to the second term in the r.h.s.
of (\ref{eq:h33}). If non--monotonicity holds, $t(h_0)$ shows plateaus at
$t_1$ and $t_2$, leading to a gap structure in the distribution of the
local fields. The resulting discontinuity in ${\rm d}h_0 / {\rm d} t$
causes a divergence of the l.h.s. of (\ref{eq:15}). This means that when
non--monotonicity holds, the RS solution is always unstable.
In the case of monotonicity, the stability condition reads
\begin{equation}
      \alpha_{RS}\left\langle\int_{-\infty}^{t_{\zeta}^{-}} {\rm D}t
            \left(\frac{1}{\gamma^2 x g''(\gamma h_0)}- 1\right)^{-2}
       +\int_{t_{\zeta}^{+}}^{\infty} {\rm D}t
             \left(\frac{1}{\gamma^2 x g''(\gamma h_0)}+1\right)^{-2}
       \right\rangle_{\{\zeta\}}<1      \,.
      \label{eq:h34}
\end{equation}

Next we consider first step RSB for the piecewise--linear input-output
relation. Similarly to the GD cost function we substitute
$t$ by $t_0\sqrt{q_0}+t_1\sqrt{1-q_0}$ in (\ref{eq:38}) in order to  obtain
$h_0(\zeta,x,q_0,t_0,t_1)$ and $F_{RSB1}(h_0,\zeta,x,q_0,t_0,t_1)$.
The free energy is obtained from (\ref{eq:22}) with the function
$\Psi(\zeta,x,q_0,M,t_0)$ (recall eq.~(\ref{eq:23})) given by
\begin{eqnarray}
 & &\Psi(\zeta,x,q_0,M,t_0) =
       \exp\left[{-M\gamma x\left(l+\frac{1}{\gamma}\right)}\right]
              \int_{-\infty}^{\Omega(h_1,q_0,t_0)}
                {\rm D}t_1\nonumber\\
       &+& \exp\left[{-M\gamma x\left(l-\frac{\gamma
                                x}{2}-t_0\sqrt{q_0}\right)}\right]
            \int_{\Omega(h_1,q_0,t_0)}^{\Omega(h_2,q_0,t_0)}{\rm D}t_1
               \exp\left[{M\gamma x t_1\sqrt{(1-q_0)}}\right]
               \nonumber\\
      & &\!\!\!\!\!\!\!\!\!\!\!\!\!\!\!\!
        +\int_{\Omega(h_2,q_0,t_0)}^{\Omega(l,q_0,t_0)}{\rm D}t_1\,
        \Phi(l,M,q_0,t_0,t_1)+\int_{\Omega(l,q_0,t_0)}^
            {\Omega(u,q_0,t_0)}{\rm D}t_1
     +\int_{\Omega(u,q_0,t_0)}^{\Omega(h_3,q_0,t_0)}{\rm D}t_1\,
        \Phi(u,M,q_0,t_0,t_1)\nonumber\\
     &+& \exp\left[{-M\gamma x\left(-u-\frac{\gamma x}{2}+t_0\sqrt{q_0}
                                 \right)}\right]
         \int_{\Omega(h_3,q_0,t_0)}^{\Omega(h_4,q_0,t_0)}{\rm D}t_1
               \exp\left[{-M\gamma x t_1\sqrt{(1-q_0)}}\right]
                         \nonumber\\
     &+& \exp\left[{-M\gamma x\left(\frac{1}{\gamma}-u\right)}\right]
          \int_{\Omega(h_4,q_0,t_0)}^{\infty}{\rm D}t_1\,.
           \label{eq:48}
\end{eqnarray}

The RSB1 $\zeta$--dependent distribution of the local fields
(\ref{eq:25}) becomes
\begin{eqnarray}
     & &\rho(h,\zeta)=\int\frac{{\rm D}t_0}{\Psi(\zeta,x,q_0,M,t_0)}
      \left\{ \frac{\exp\left[-M\gamma x\left(l+\frac{1}{\gamma}\right)
          -\frac{1}{2}\Omega^2(h,q_0,t_0)\right]}
       {\sqrt{2\pi(1-q_0)}}\theta\left(h_1-h\right)\right.
       \nonumber\\
    &+&\frac{\exp\left[-M\gamma x\left(l+\frac{\gamma x} {2}-h\right)
        -\frac{1}{2}\Omega^2(h-\gamma x,q_0,t_0)\right]}
           {\sqrt{2\pi(1-q_0)}}\left[\theta(h-h'_2)- \theta(h-l)\right]
        \theta\left(l+\frac{1}{\gamma} - \frac{\gamma x}{2} \right)
           \nonumber\\
   &+&\delta(h-l)\int_{\Omega(h_2,q_0,t_0)}^{\Omega(l,q_0,t_0)}
        {\rm D}t_1\,\Phi(l,M,q_0,t_0,t_1)
    +\frac{\exp\left[-\frac{1}{2}\Omega^2(h,q_0,t_0)\right]}
        {\sqrt{2\pi (1-q_0)}}\left[\theta(h-l)-\theta(h-u)\right]
            \nonumber\\
    & &\quad\quad+\delta(h-u)\int_{\Omega(u,q_0,t_0)}^{\Omega((h_3,q_0,t_0)}
        {\rm D}t_1\,\Phi(u,M,q_0,t_0,t_1)
            \nonumber\\
   &+&\frac{\exp\left[-M\gamma x\left[h-u+\frac{\gamma x}{2}\right]
        -\frac{1}{2}\Omega^2(h+\gamma x,q_0,t_0)\right]}
         {\sqrt{2\pi (1-q_0)}}\left[\theta(h-u)-\theta(h-h'_3)\right]
        \theta\left(\frac{1}{\gamma} - \frac{\gamma x}{2} -u \right)
            \nonumber\\
  & &\quad+\left.\frac{\exp\left[-M\gamma x\left(\frac{1}{\gamma}-u\right)
           -\frac{1}{2}\Omega^2(h,q_0,t_0)\right]}
                        {\sqrt{2\pi(1-q_0)}}\theta(h-h_4)\right\}\,.
        \label{eq:49}
\end{eqnarray}

Finally, the $\zeta$--dependent RSB1 average output error reads
\begin{eqnarray}
  & &{\cal E}(\zeta) =\int\frac{{\rm D}t_0}{\Psi(\zeta,x,q_0,M,t_0)}
         \frac{\gamma}{\sqrt{2\pi(1-q_0)}}
             \nonumber\\
  & &\times\left\{\left(l+\frac{1}{\gamma}\right)
          \exp\left[{-M\gamma x\left(l+\frac{1}{\gamma}\right)}\right]
              \int_{-\infty}^{h_1}
              {\rm d}h\,\exp\left[-
\frac{1}{2}\Omega^2(h,q_0,t_0)\right]\right.
                \nonumber\\
   &+&\int_{h'_2}^l{\rm d}h\,(l-h)\exp\left[-\frac{1}{2}\Omega^2(h-
        \gamma x,q_0,t_0)-M\gamma
               x\left(l+\frac{\gamma x} {2}-h\right)\right]
        \theta\left(l+\frac{1}{\gamma} - \frac{\gamma x}{2} \right)
                 \nonumber\\
    &+&\int_u^{h'_3}{\rm d}h\,(h-u)\exp\left[-\frac{1}{2}\Omega^2(h
        +\gamma x,q_0,t_0)-M\gamma x\left(h-u+
             \frac{\gamma x} {2}\right)\right]
        \theta\left(\frac{1}{\gamma} - \frac{\gamma x}{2} -u \right)
             \nonumber\\
    &+&\left.\left(\frac{1}{\gamma}-u\right)
       \exp\left[{-M\gamma x\left(\frac{1}{\gamma}-u\right)}\right]
               \int_{h_4}^{\infty}{\rm d}h\,
          \exp\left[-\frac{1}{2}\Omega^2(h,q_0,t_0)\right]\right\}\,.
        \label{eq:50}
\end{eqnarray}

\subsection{Quadratic cost function}
\pin
Again we look at the RS treatment first. We start by defining
\begin{eqnarray}
      &&  h_1=l-\sqrt{1+2\gamma^2 x}\left(l+\frac{1}{\gamma}\right)\,,
        \label{eq:51} \\
      &&  h_2=l-\frac{l+\frac{1}{\gamma}}{\sqrt{1+2\gamma^2 x}}\,,
        \label{eq:52} \\
      &&  h_3=u+\frac{\frac{1}{\gamma}-u}{\sqrt{1+2\gamma^2 x}}\,,
        \label{eq:53} \\
      &&  h_4=u+\sqrt{1+2\gamma^2 x}\left(\frac{1}{\gamma}-u\right)\,.
        \label{eq:54}
\end{eqnarray}
Using these definitions, the result of the minimization in $h$ of $F_{RS}$
(Eq.~(\ref{eq:13})) becomes
\begin{eqnarray}
    \begin{array}{lll}
    h_0=t\,, & \quad  F_{RS}(h_0,\zeta,x,t)=\gamma^2\left(l+\frac{1}
              {\gamma}\right)^2 & \mbox{for} \quad -\infty<t<h_1 \\
    h_0=\frac{2\gamma^2 xl+t}{1+2\gamma^2 x}\,, & \quad
            F_{RS}(h_0,\zeta,x,t)=\gamma^2\frac{(l-t)^2}{1+2\gamma^2 x}
            & \mbox{for} \quad h_1<t<l\\
    h_0=t\,,&\quad  F_{RS}(h_0,\zeta,x,t)=0 & \mbox{for} \quad  l<t<u \\
    h_0=\frac{2\gamma^2 xu+t}{1+2\gamma^2 x}\,, & \quad
         F_{RS}(h_0,\zeta,x,t)=\gamma^2\frac{(u-t)^2}{1+2\gamma^2 x}
                & \mbox{for} \quad u<t<h_4\\
    h_0=t\,, & \quad  F_{RS}(h_0,\zeta,x,t)=\gamma^2\left(\frac{1}
               {\gamma}-u\right)^2 & \mbox{for} \quad h_4<t<\infty \\
        \end{array}
        \label{eq:55}
\end{eqnarray}
The RS saddle--point equation is obtained from (\ref{eq:14}) and
(\ref{eq:55}):
\begin{equation}
        \alpha_{RS}^{-1}=\left(\frac{2\gamma^2 x}{1+2\gamma^2 x}\right)
                       \left\langle
        \int_{h_1}^l {\rm D}t(t-l)^2+\int_u^{h^4} {\rm D}t (t-u)^2
                      \right\rangle_{\zeta}\,.
              \label{eq:56}
\end{equation}
>From (\ref{eq:19}) and (\ref{eq:55}), the RS $\zeta$--dependent
distribution of the local fields becomes
\begin{eqnarray}
    &&\rho(h,\zeta)
    = \frac{\exp\left[{-\frac{h^2}{2}}\right]}{\sqrt{2 \pi}}
              \left[\theta(h_1-h)
             +\theta(h-l)-\theta(h-u)+\theta(h-h4)\right]
             \nonumber\\
   &+& \left(1+2\gamma^2 x\right)\left\{
      \frac{\exp\left[{-\frac{1}{2} \left(\left(1 +2\gamma^2 x\right)h
               -2\gamma^2 xl\right)^2}\right]}{\sqrt{2\pi}}
          \left[\theta(h-h_2)-\theta(h-l)\right]\right.
           \nonumber\\
   &+&\left.\frac{\exp\left[{-\frac{1}{2}\left(\left(1+2\gamma^2 x
             \right)h-2\gamma^2 xu\right)^2}\right]}{\sqrt{2\pi}}
             \left[\theta(h-hu)-\theta(h-h_3) \right]\right\}\,.
           \label{eq:57}
\end{eqnarray}
Consequently, the RS $\zeta$--dependent output error becomes
\begin{eqnarray}
  {\cal E}(\zeta) &=& \gamma \left[\left(l+\frac{1}{\gamma}\right)
            \int_{-\infty}^{h_1}\frac{{\rm d}h}{\sqrt{2\pi}}\exp\left[
                         {-\frac{h^2}{2}}\right]\right.
                         \nonumber\\
     &+& \left. \left(1+2\gamma^2 x\right)\int_{h_2}^l{{\rm d}h}
                        {\sqrt{2\pi}}
      \exp\left[{\frac{\left(\left(1+2\gamma^2 x\right)h-2\gamma^2
               xl \right)^2}{2}}\right](l-h) \right.
              \nonumber\\
    &+& \left.\left(1+2\gamma^2 x\right)\int_u^{h_3}{{\rm d}h}
                             {\sqrt{2\pi}}
       \exp\left[{-\frac{\left(\left(1+2\gamma^2 x\right)h-
                     2\gamma^2 xu\right)^2}
              {2}}\right](h-u)\right.
              \nonumber\\
    &+& \left.\left(\frac{1}{\gamma}-u\right)\int_{h_4}^
                  {-\infty}\frac{{\rm d}h} {\sqrt{2\pi}}
              \exp\left[{-\frac{h^2}{2}}\right]\right]\,.
        \label{eq:58}
\end{eqnarray}

For the hyperbolic tangent input--output relation, $h_{\zeta}$ (recall
eq.~(\ref{eq:17})) is no longer a local minimum of (\ref{eq:13}). The
minima are always given by the solutions of (\ref{eq:h30}), that define
$t(h_0)$ as
\begin{equation}
          t(h_0)=h_0+2\gamma x g'(\gamma h_0)\left[g(\gamma h_0)
                 -\zeta\right]  \,.
           \label{eq:h35}
\end{equation}
Depending on the values of $\gamma^2 x$ {\it and} $\zeta$, $t(h_0)$ is a
monotonic function or not. The onset to non--monotonicity is given by the
system of equations (\ref{eq:h32}). If monotonicity holds, $h_0(t)$ is
continuous, and is obtained by inverting (\ref{eq:h35}). Otherwise,
$h_0(t)$ displays one or two jumps at $t=t_1$ and/or $t=t_2$, whose values
are obtained by a Maxwell construction.

The distribution of local fields is obtained directly from (\ref{eq:19})
\begin{equation}
    \rho(h|\zeta)=\frac{{\rm d}t}{{\rm d}h}
         \frac{\exp\left[{-\frac{t^2 (h)}{2}}\right]}{\sqrt{2\pi}}   \,.
      \label{eq:h36}
\end{equation}
When non--monotonicity holds, the jumps in $h_0(t)$ give rise to a gap
structure in the distribution of the local fields and the RS solution
becomes unstable. In the monotonic case the stability condition
(\ref{eq:15}) reads
\begin{equation}
   \alpha_{RS}\left\langle\int_{-\infty}^{+\infty} {\rm D}t
                \left(\frac{1}{2\gamma^2 x \left[\left(g'(\gamma
          h_0)\right)^2+\left(g(\gamma h_0)-\zeta\right)
            g''(\gamma h_0)\right]}+1\right)^{-2}
               \right\rangle_{\{\zeta\}}<1      \,.
      \label{eq:h37}
\end{equation}

Let us finally turn to the RSB1 treatment. Again, in order to obtain
$h_0(\zeta,x,q_0,t_0,t_1)$ and $F_{RSB1}(h_0, \zeta,x,q_0,t_0,t_1)$
we substitute $t$ by $t_0\sqrt{q_0}+t_1\sqrt{1-q_0}$ in (\ref{eq:38}).
The free energy is obtained from (\ref{eq:22}), whereby the function
$\Psi(\zeta,x,q_0,M,t_0)$, given by (\ref{eq:23}), becomes
\begin{eqnarray}
    & &\Psi(\zeta,x,q_0,M,t_0) =
         \exp\left[{-M\gamma^2 x\left(l+\frac{1}{\gamma}\right)^2}\right]
                \int_{-\infty}^{\Omega(h_1,q_0,t_0)}{\rm D}t_1
                      \nonumber\\
  &+&\int_{\Omega(h_1,q_0,t_0)}^{h_2,q_0,t_0)}{\rm D}t_1
     \Phi(l,\frac{2M\gamma^2 x}{1+2\gamma^2 x},q_0,t_0)
   +\int_{\Omega(l,q_0,t_0)}^{\Omega(u,q_0,t_0)}{\rm D}t_1\nonumber\\
   &+&\int_{\Omega(u,q_0,t_0)}^{h_4,q_0,t_0)}{\rm D}t_1
     \Phi(u,\frac{2M\gamma^2 x}{1+2\gamma^2 x},q_0,t_0)
        +\exp\left[{-M\gamma^2 x\left(\frac{1}{\gamma}-u\right)^2}\right]
         \int_{\Omega(u,q_0,t_0)}^{\infty}{\rm D}t_1\nonumber\\
\label{eq:59}
\end{eqnarray}

For the RSB1 $\zeta$--dependent distribution of the local fields we
obtain
\begin{eqnarray}
    \rho(h,\zeta) &=&\int\frac{{\rm D}t_0}{\Psi(\zeta,x,q_0,M,t_0)}
    \left\{ \frac{\exp\left[-M\gamma x\left(l+\frac{1}{\gamma}\right)
        -\frac{1}{2}\Omega^2(h,q_0,t_0) \right]}
         {\sqrt{2\pi(1-q_0)}}\theta\left(h_1-h\right)\right.
              \nonumber\\
    &+&\frac{\exp\left[-M\gamma x\left(l+\frac{\gamma x} {2}-h\right)
      -\frac{1}{2}\Omega^2(h-\gamma x, q_0,t_0)\right]}
          {\sqrt{2\pi(1-q_0)}}\left[\theta(h-h'_2)- \theta(h-l)\right]
          \nonumber\\
    &+&\delta(h-l)\int_{\Omega(h_2,q_0,t_0)}^{\Omega(l,q_0,t_0)}{\rm D}t_1
                \Phi(M,l,q_0,t_0,t_1)\nonumber\\
  & &\quad +\frac{\exp\left[-\frac{1}{2}\Omega(h,q_0,t_0)\right]}
        {\sqrt{2\pi (1-q_0)}}\left[\theta(h-l)-\theta(h-u)\right]
             \nonumber\\
   & &\quad +\delta(h-u)\int_{\Omega(u,q_0,t_0)}^{\Omega(h_3,q_0,t_0}{\rm 
D}t_1
        \Phi(M,u,q_0,t_0,t_1)
              \nonumber\\
  &+&\frac{\exp\left[-M\gamma x\left[h-u+\frac{\gamma x}{2} \right]
       - \frac{1}{2}\Omega(h+\gamma x,q_0,t_0)\right]}
          {\sqrt{2\pi (1-q_0)}}\left[\theta(h-u)-\theta(h-h'_3) \right]
          \nonumber\\
 &+&\left.\frac{\exp\left[-M\gamma x\left(\frac{1}{\gamma}- u\right)
        -\frac{1}{2}\Omega(h,q_0,t_0)\right]}
           {\sqrt{2\pi(1-q_0)}}\theta(h-h_4)\right\}\,.
        \label{eq:60}
\end{eqnarray}
The $\zeta$--dependent RSB1 average output error becomes
\begin{eqnarray}
  & &{\cal E}(\zeta) =\int\frac{{\rm D}t_0}{\Psi(\zeta,x,q_0,M,t_0)}
         \frac{\gamma}{\sqrt{2\pi(1-q_0)}}
             \nonumber\\
  & &\times\left\{\left(l+\frac{1}{\gamma}\right)
          \exp\left[{-M\gamma^2 x\left(l+\frac{1}{\gamma}\right)^2}\right]
              \int_{-\infty}^{h_1}
              {\rm d}h\,\exp\left[-
\frac{1}{2}\Omega^2(h,q_0,t_0)\right]\right.
                \nonumber\\
   &+&(1+2\gamma^2 x)\int_{h_2}^l{\rm d}h\,(l-h)
        \exp\Bigl[-\frac{1}{2}\Omega^2((1+2\gamma^2 x)(h-l)+l,q_0,t_0)
        \nonumber\\
        & &\quad\quad -M\gamma^2 x(1+2\gamma^2 x)(h-l)^2\Bigr]\nonumber\\
             \nonumber\\
   &+&(1+2\gamma^2 x)\int_u^{h_3}{\rm d}h\,(h-u)
        \exp\Bigl[-\frac{1}{2}\Omega^2((1+2\gamma^2 x)(h-u)+u,q_0,t_0)
        \nonumber\\
    & &\quad\quad -M\gamma^2 x(1+2\gamma^2 x)(h-u)^2\Bigr]\nonumber\\
    &+&\left.\left(\frac{1}{\gamma}-u\right)
       \exp\left[{-M\gamma^2 x\left(\frac{1}{\gamma}-u\right)^2}\right]
               \int_{h_4}^{\infty}{\rm d}h\,
          \exp\left[-\frac{1}{2}\Omega^2(h,q_0,t_0)\right]\right\}\,.
        \nonumber \\
        \label{eq:61}
\end{eqnarray}

\newpage
\section*{Figure Captions}
\noindent
{\bf Figure 1} : Storage capacity $\alpha$ for the hyperbolic tangent
input--output relation as a function of the gain parameter $\gamma$ at
constant total average output error ${\cal E}= 0$ (lower
curve), $0.1$, $0.2$ and $0.4$ (upper curve) for the GD cost function (a),
the linear cost function (b) and the quadratic cost function (c).
In (b) and (c) the line for ${\cal E}=0.4$ is not shown. The dotted curve
is the AT--line.\\
\noindent
{\bf Figure 2} : The gap structure for the hyperbolic tangent input--output
relation in the $\alpha-\gamma $--plane for the
linear cost function (a) and the quadratic cost function (b). The
curve $\alpha_c$ is the critical capacity, the curve $\alpha_g$ represents
the gap line and $\alpha_{AT}$ is the AT--line.\\
\noindent
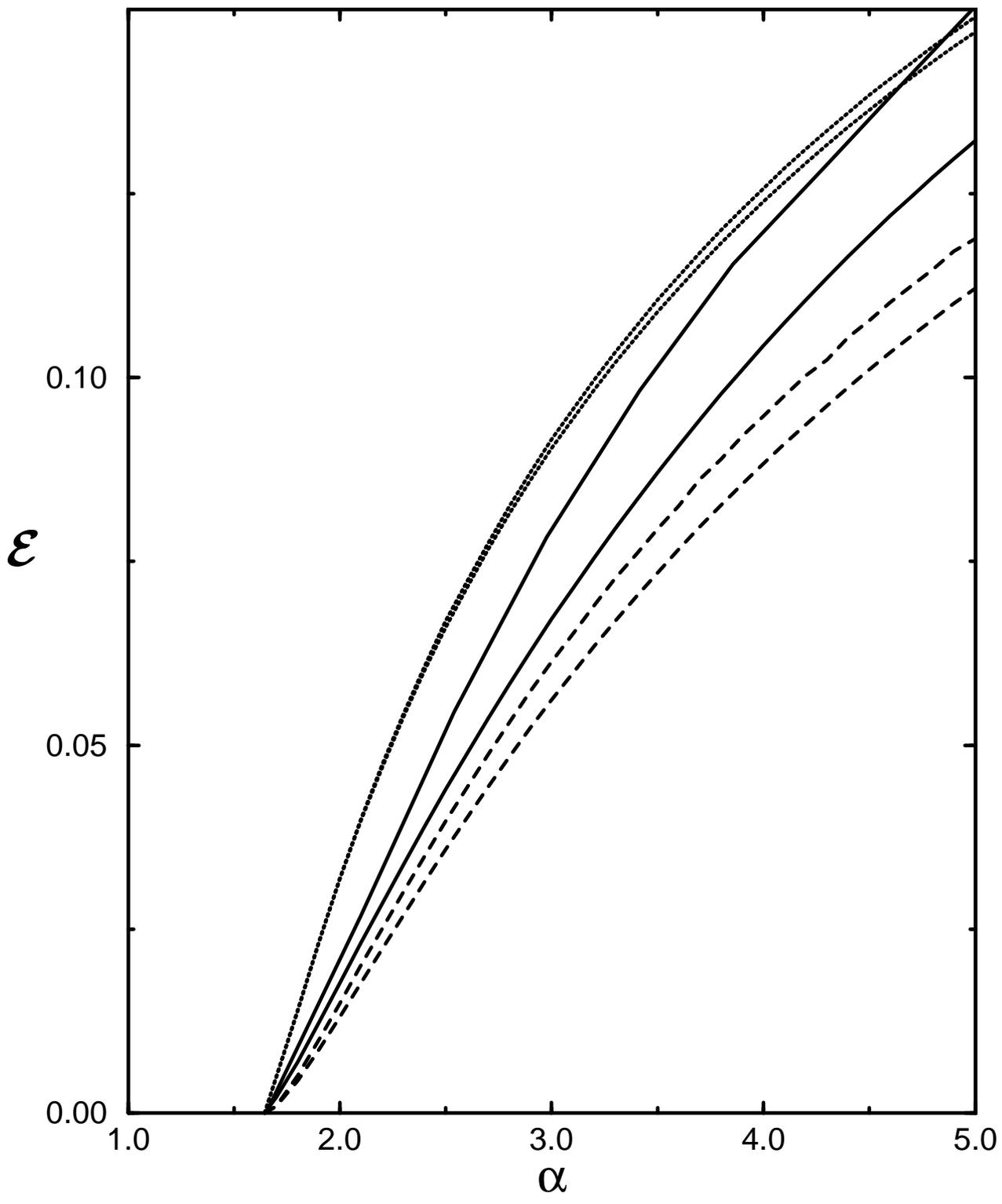
{\bf Figure 3} : The total average output error ${\cal E}$ as a function of
the storage capacity $\alpha $ for output tolerance $\epsilon=0.5$ and gain
parameter $\gamma=1$ in the case of the piecewise linear input--output
relation for the GD cost function (solid lines), the linear cost
function (dashed curves) and the quadratic cost function (dotted lines).
For each case the upper curve is the RSB1 result, the lower one the RS
result.\\
\noindent
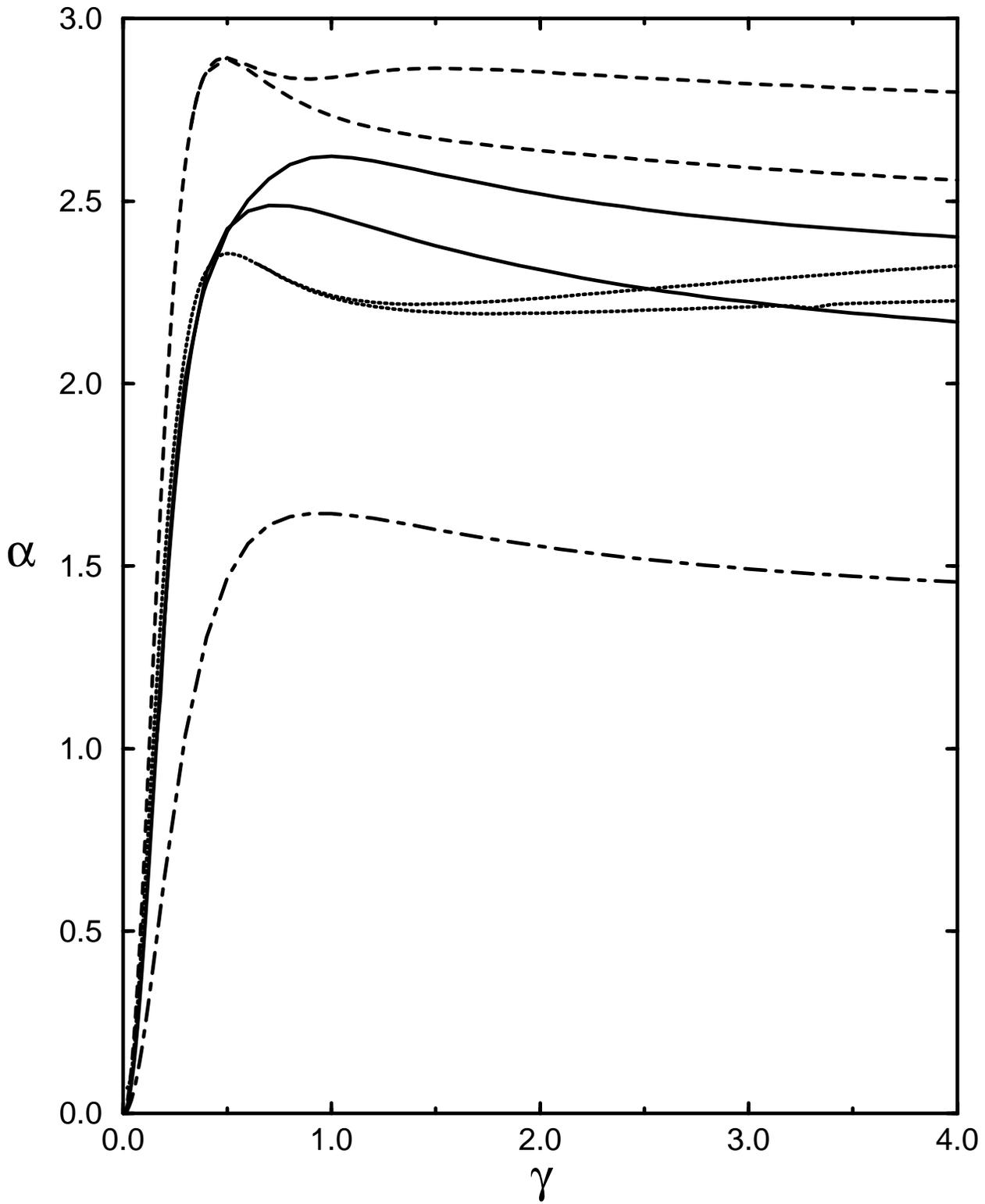
{\bf Figure 4} : Storage capacity $\alpha$ for the piecewise linear
input--output relation as a function of the gain parameter $\gamma$ for an
output tolerance $\epsilon=0.5$ and a total average output error ${\cal E}=
0.05$ for the GD cost function (solid lines), the linear cost function
(dashed curves) and the quadratic cost function (dotted curves). For each
case the upper curve is the RS result, the lower one the RSB1 result.
The dashed--dotted curve is the critical storage capacity.\\
\noindent
{\bf Figure 5} : Distribution of the local fields for the piecewise linear
input--output relation and $\alpha=3, \gamma=1, \epsilon=0.5$ and the
correct output $\zeta=0.6$ for the GD cost function (a), the linear
cost function (b) and the quadratic cost function (c). In (c) the results
on
the interval $h=[-3,-1]$ are magnified by a factor $20$. The dotted curves
are the RS results, the solid lines the RSB1 results.\\

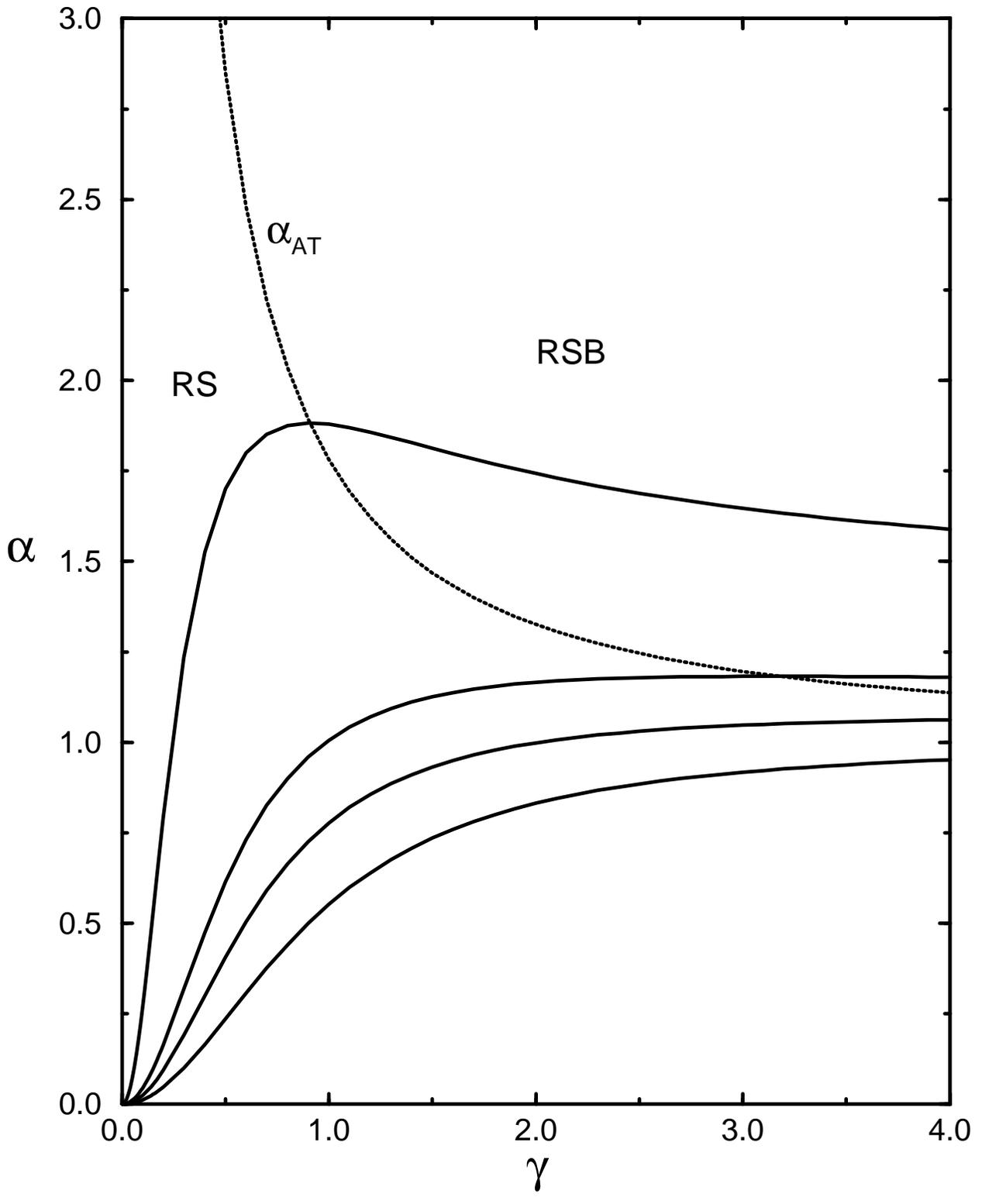
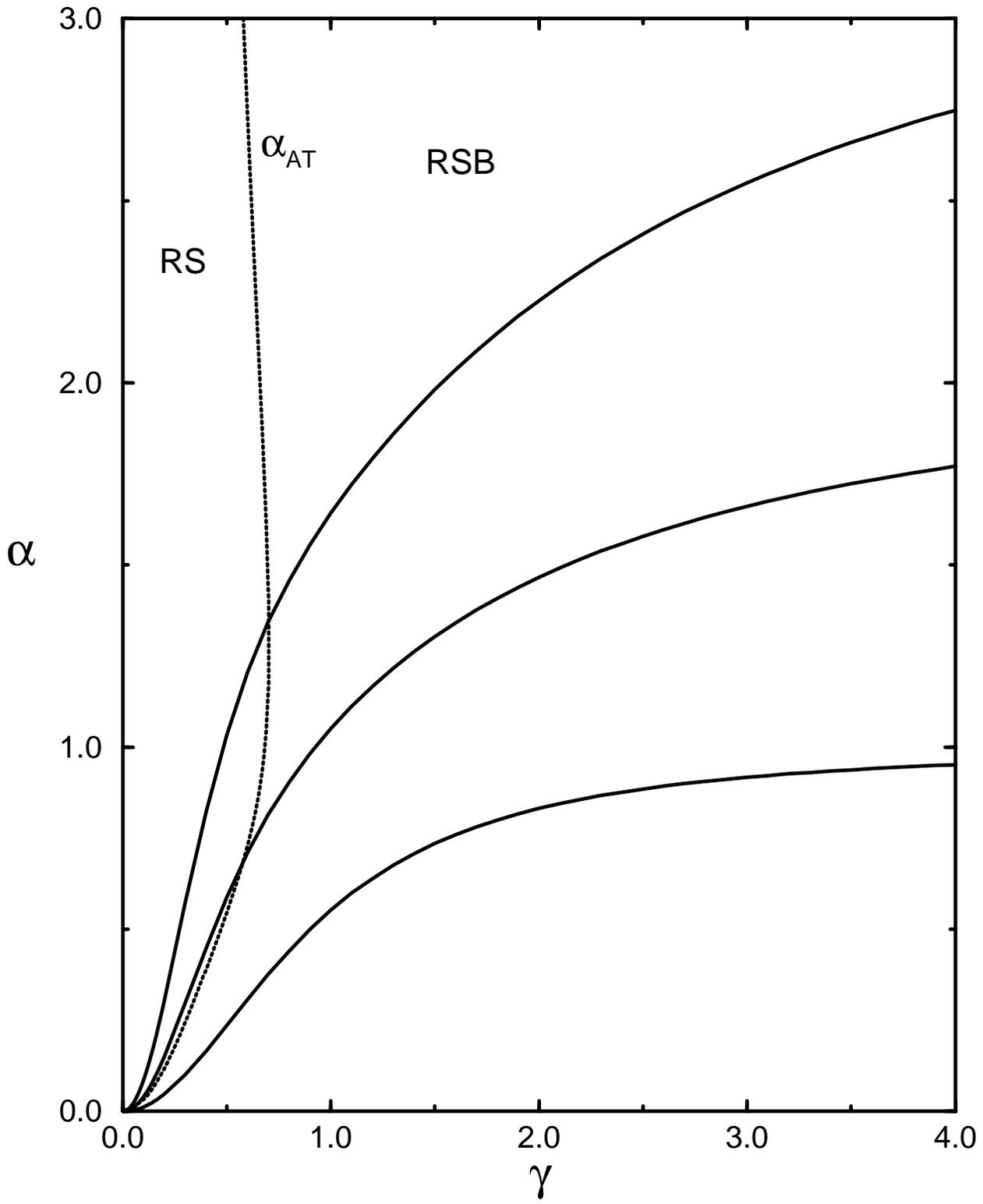
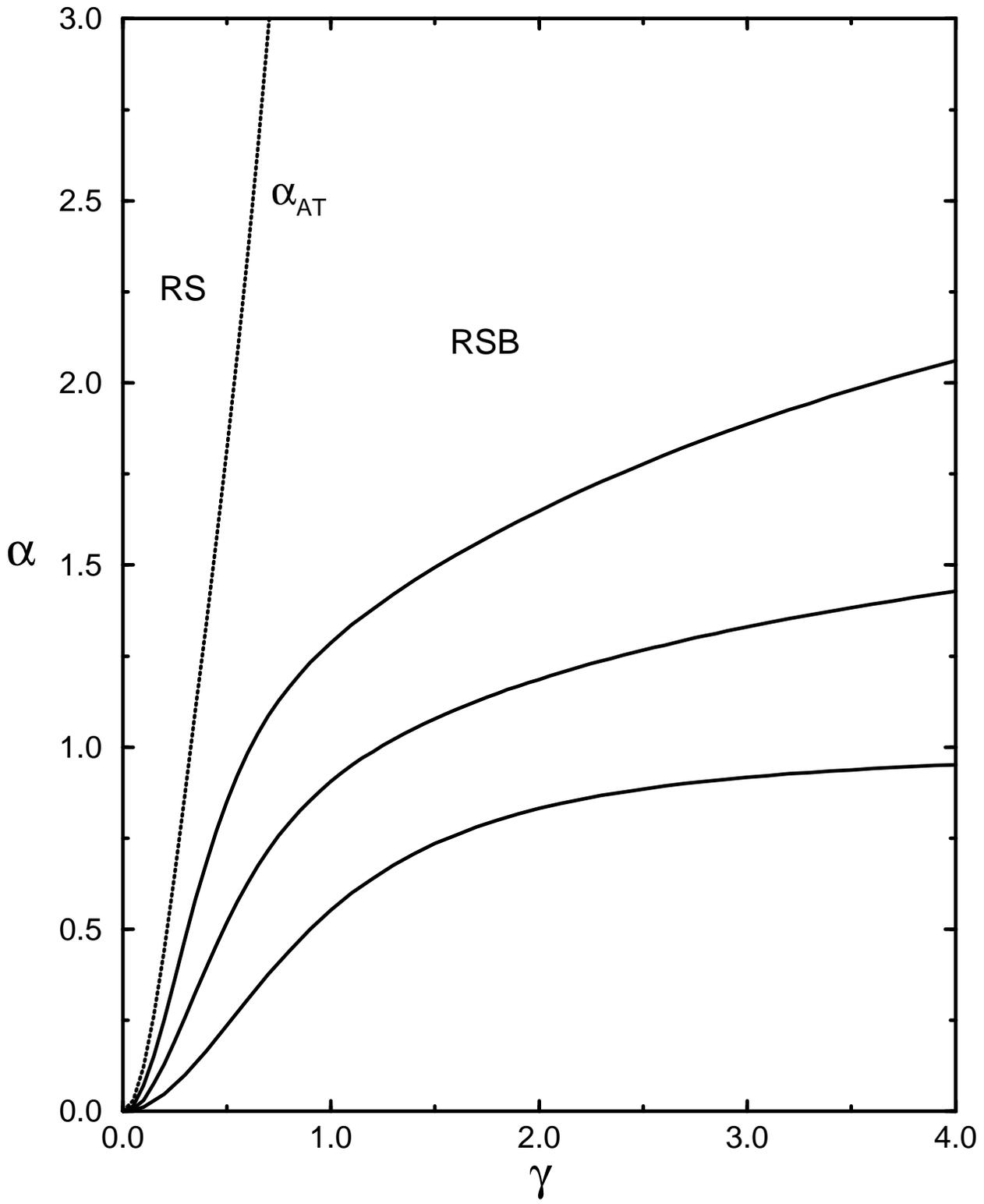
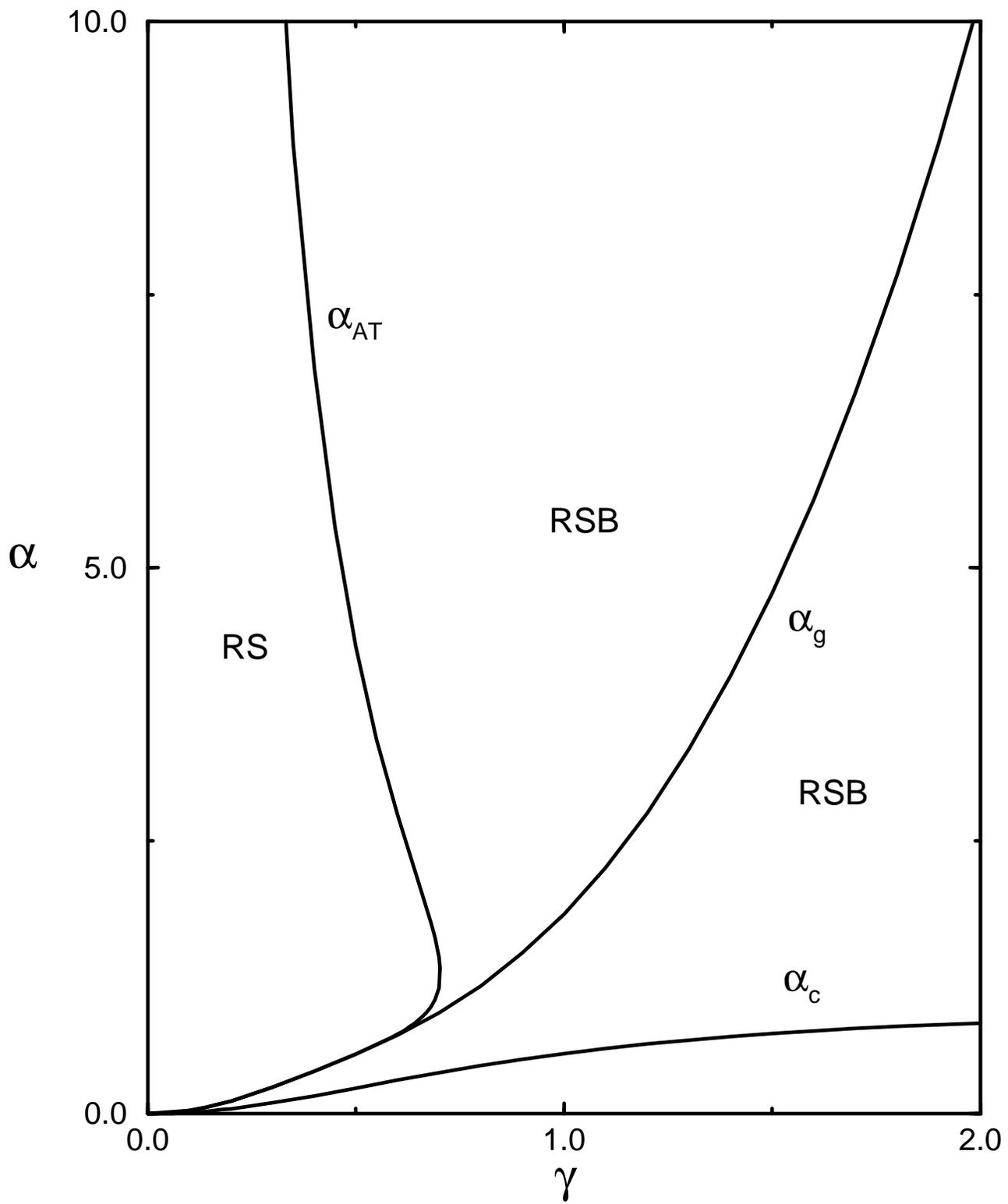
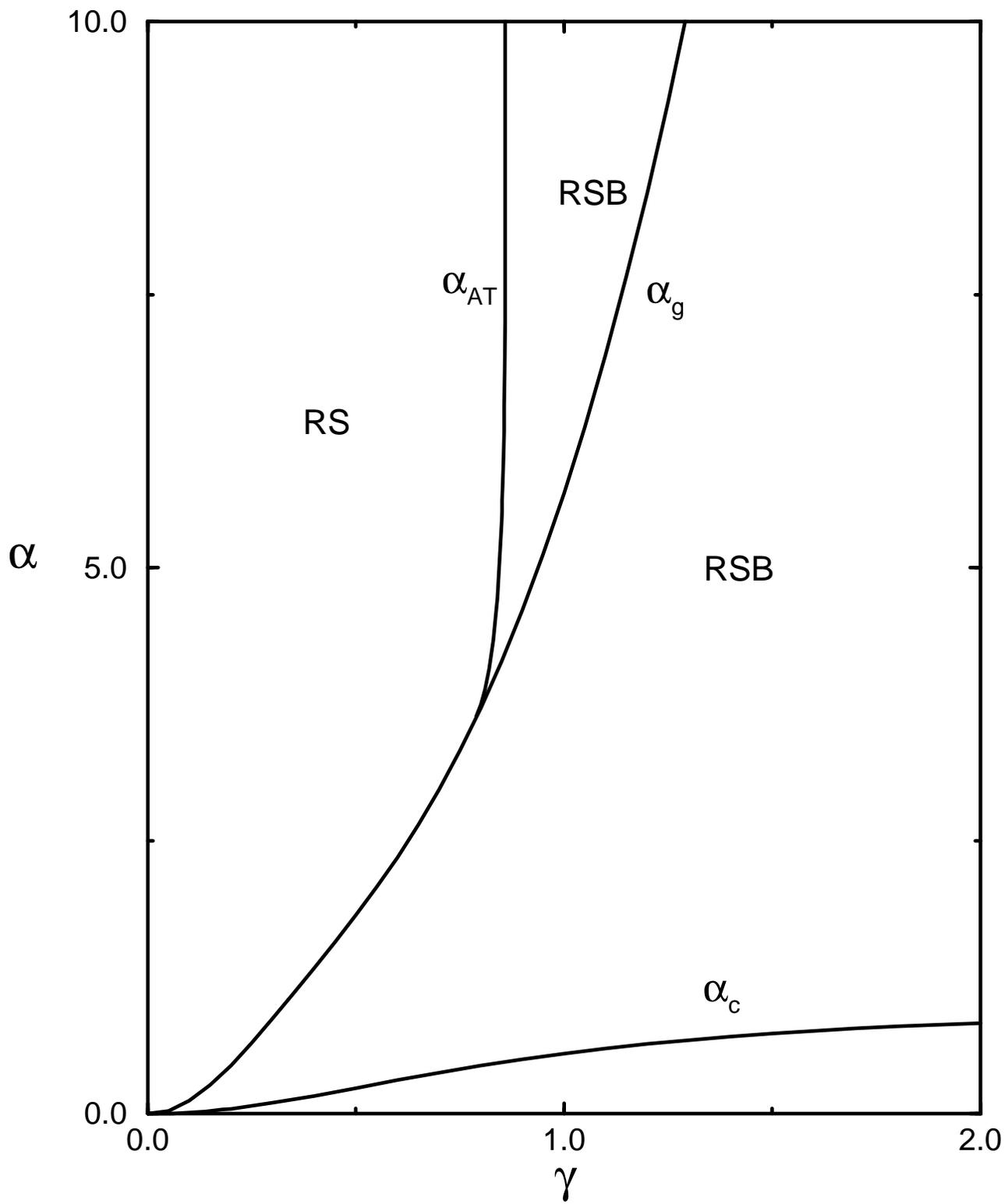
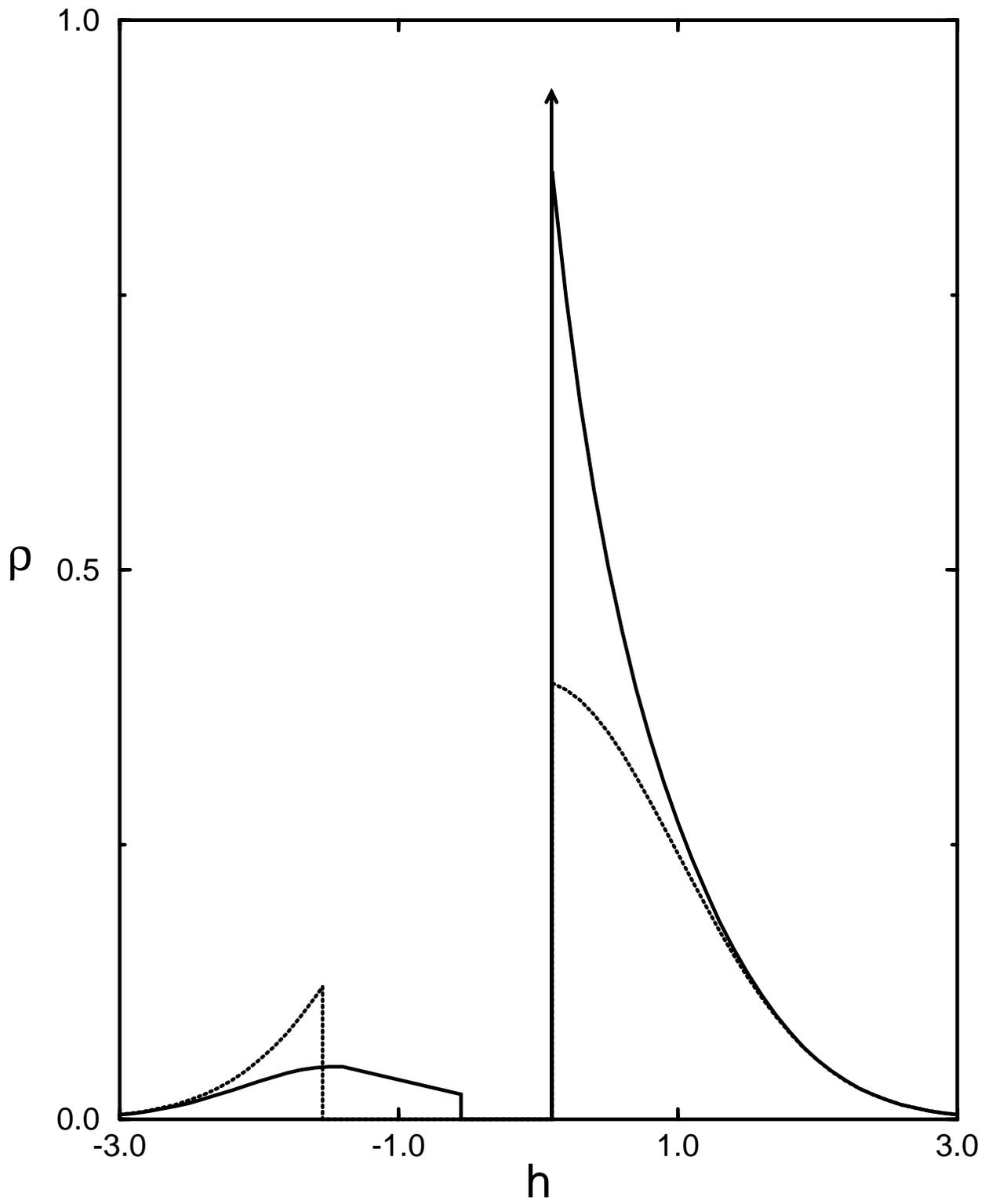
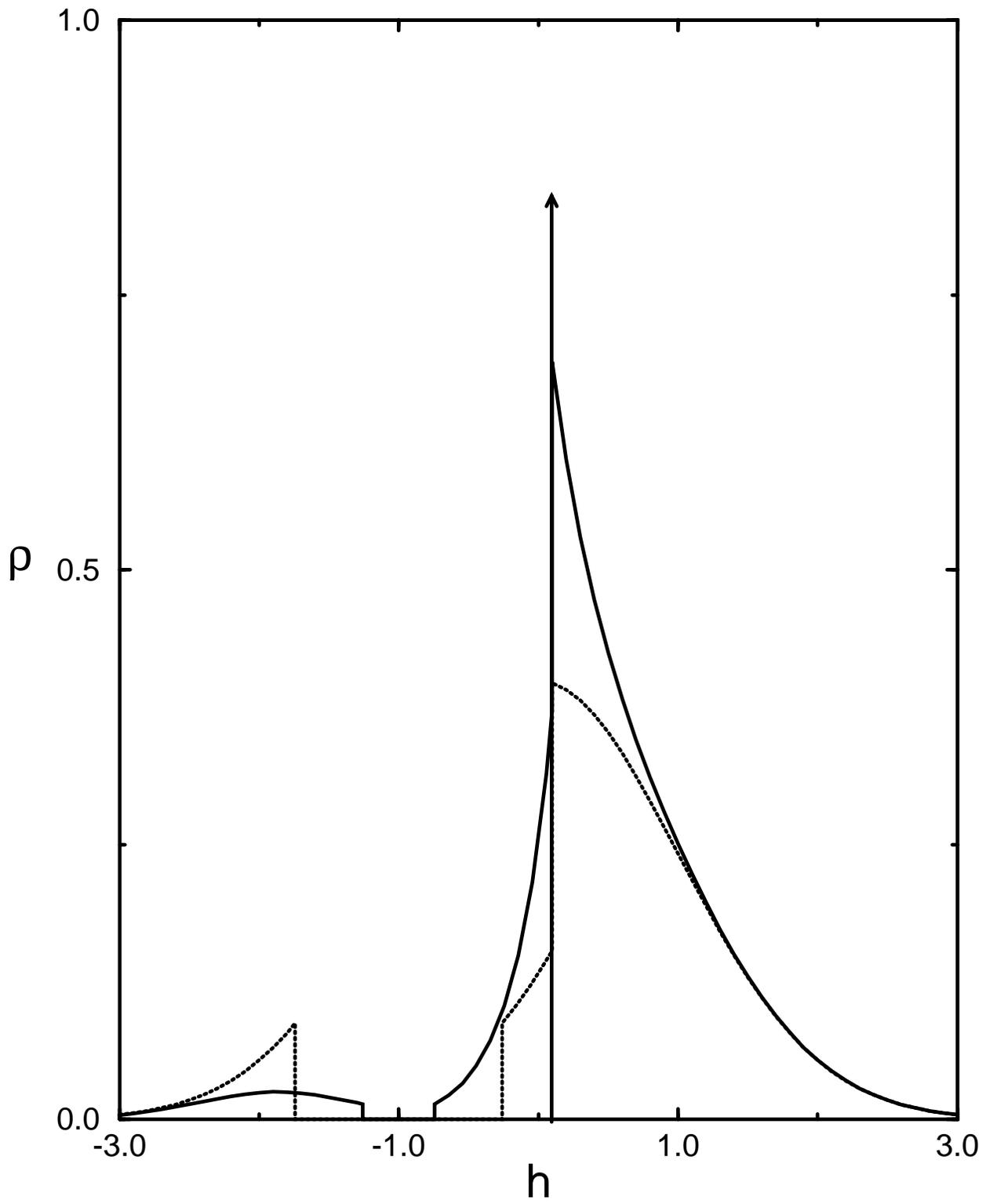
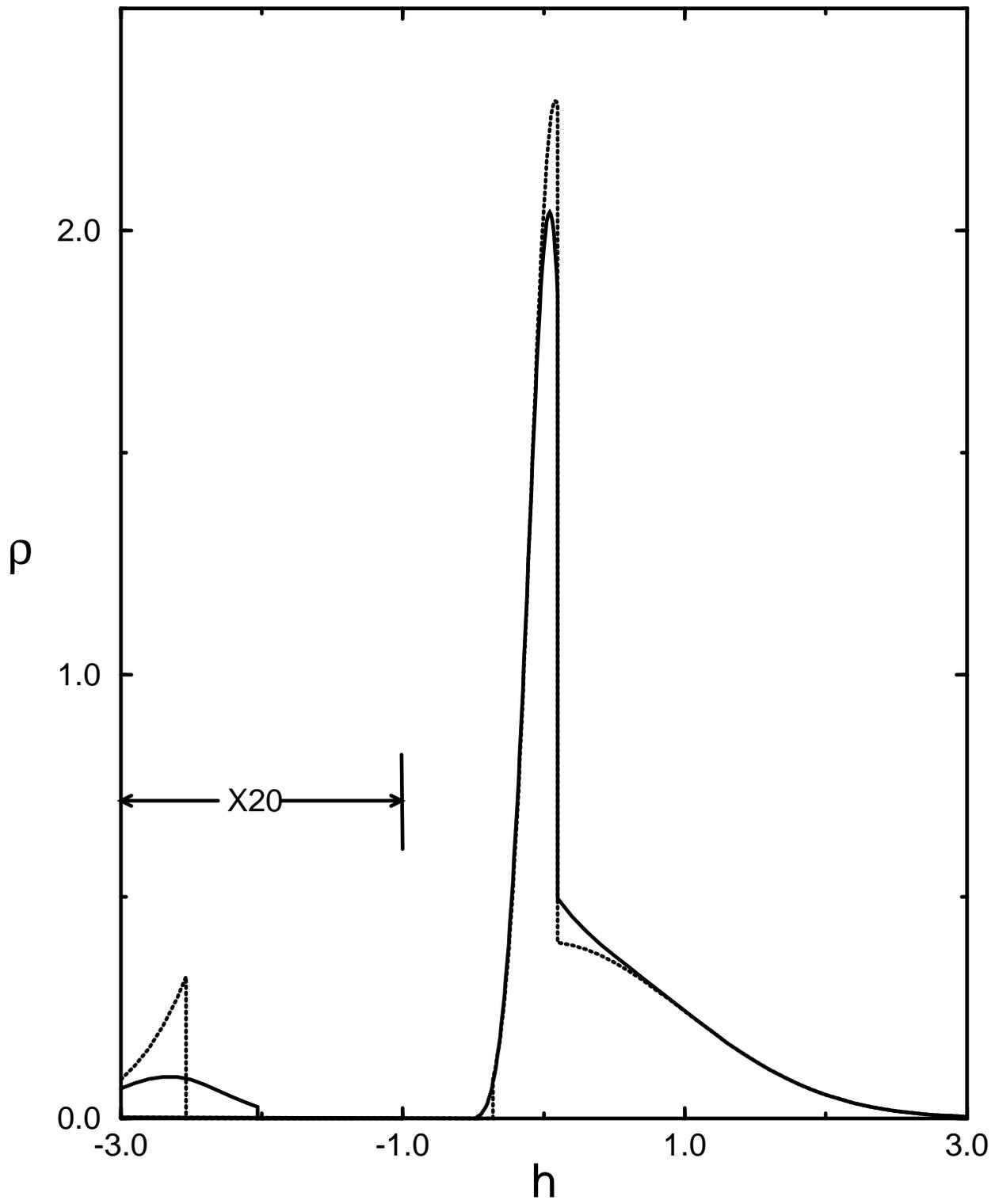
\end{document}